\documentclass[usegraphicx,usedcolumn]{mn2e}
\usepackage{color,epsfig,rotating,amssymb,longtable}
\usepackage{array,colortbl,lscape}

\newif\ifAMStwofonts

\usepackage{color}

\definecolor{dgreen}{rgb}{0,.5,.1} 
\definecolor{pink}{rgb}{.9,.2,.5}  
\definecolor{orange}{rgb}{.9,.4,0} 
\definecolor{darkred}{rgb}{.545,0.0,.0}
\begin{document}

\title[Tailor-made models of HII galaxies]{Disentangling the metallicity and star formation history of HII galaxies through tailor-made models }
\author[E. P{\'e}rez-Montero et al.]
       {Enrique P{\'e}rez-Montero$^{1,2}$, Rub\'en Garc\'{\i}a-Benito$^{3}$,
Guillermo F. H\"agele$^{3,4}$
\newauthor \& \'Angeles I. D\'{\i}az$^{3}$\\
$^{1}$ Instituto de Astrof\'\i sica de Andaluc\'\i a, CSIC, Apdo. 3004, 18080, Granada, Spain.\\
$^{2}$ Laboratoire d'Astrophysique de Toulouse-Tarbes. Observatoire Midi-Pyr\'en\'ees-CNRS-Universit\'e de Toulouse 3,\\
 14, avenue Edouard Belin. 31400. Toulouse. France\\  
 $^{3}$ Departamento de F\'{\i}sica Te\'orica, C-XI, Universidad Aut\'onoma de
Madrid, 28049 Madrid, Spain\\ 
$^{4}$ Facultad de Cs Astron\'omicas y Geof\'isicas, Universidad Nacional de La
Plata, Paseo del Bosque s/n, 1900 La Plata, Argentina \\
 }

\date{Accepted 
      Received ;
      in original form January 2010}

\pagerange{\pageref{firstpage}--\pageref{lastpage}}
\pubyear{2009}

\maketitle

\label{firstpage}

\begin{abstract}
We present a self-consistent study of the stellar populations and the ionized gas in
a sample of 10 HII galaxies with, at least, four measured electron temperatures and
a precise determination of ionic abundances following the "direct method".
We fitted the spectral energy distribution of the galaxies using the program STARLIGHT
and Starburst99 libraries in order to quantify the contribution of the underlying stellar
population to the equivalent width of H$\beta$ (EW(H$\beta$)), which amounts to about 10~\% 
for most of the objects. We
then studied the Wolf-Rayet stellar populations detected in seven of the galaxies. The
presence of these populations and the EW(H$\beta$) values, once corrected for the 
continuum contribution from underlying stars and UV dust absorption, indicate that the ionizing
stellar populations were created following a continuous star formation episode of 10 Myr duration,
hence WR stars may be present in all of objects even
if they are not detected in some of them.

The derived stellar features, the number of ionizing photons
and the relative intensities of the most prominent emission lines were used as input
parameters to compute tailored models with the photoionization code CLOUDY. Our
models are able to adequately reproduce the thermal and ionization structure of these
galaxies as deduced from their collisionally excited emission lines. This indicates
that ionic abundances can be derived following the "direct method" if the thermal structure
of the ionized gas is well traced, hence no abundance discrepancy factors are
implied for this kind of objects. Only the electron temperature of S$^+$ is 
overestimated by the models, with the corresponding underestimate of its abundance, 
pointing to the possible presence of outer shells
of diffuse gas in these objects that have not been taken into account in our models.
This kind of geometrical effects can affect the determination of the equivalent effective
temperature of the ionizing cluster using calibrators which depend on low-excitation
emission lines.

\end{abstract}


\section{Introduction}

The understanding of the physical properties of gas, stars, and dust and their 
interrelations and evolution in different scenarios of star formation
constitute one of the most intriguing open issues on astrophysics. 
This type of phenomenon is well studied at different spatial scales,
since massive young star clusters are powerful sources of radiation, 
detectable even in the most distant galaxies.
For instance, from a statistical point of view, large surveys of starburst galaxies have 
been used to confirm, with a great confidence base, some of the observational 
relations between the physical properties of different types of galaxies at different 
cosmological epochs.
This is the case of the luminosity-metallicity relation 
or its equivalent, the mass-metallicity relation, which establishes that
in the local Universe the most massive galaxies 
have, in average, higher metallicities than the less massive ones 
(SDSS: Tremonti et al., 2004; 2dFGRS: Lamareille et al., 2004). 
This observational fact is  a consequence of higher stellar yields in 
the more massive galaxies, which could find less troubles to retain the metals 
ejected by stars and pushed out towards the interstellar and intergalactic medium 
by supernovae and stellar winds (Larson, 1974).

Nevertheless, since this kind of studies are mostly based on the relation 
between the stellar mass and the oxygen gas-phase abundance, they present 
some drawbacks, involving several aspects
in the determination of their properties, including the following:
1) The relation between the total mass of a galaxy and its
stellar mass depends strongly on different environmental and morphological 
circumstances (Bell \& de Jong, 2001).
2) The stellar mass is usually estimated by means of stellar synthesis population 
model fitting of the spectral energy distribution (SED). These models are often affected by
degeneracies between age, metallicity, and dust-extinction which are never well solved 
(Renzini \& Buzzoni, 1986).
3) The metallicity is only estimated in starburst galaxies through the oxygen gas phase 
abundance derived from the relative intensities of the most prominent emission lines. 
This excludes from these studies the most quiescent and early-type galaxies.
4) The spatial distribution of metallicity in massive, non-compact objects is often 
neglected, assigning a unique value to the whole distribution. 
However, it is known that the radial profile of the oxygen abundance in some
close spiral galaxies can spread over more than an order of magnitude (Searle, 1971).
5) The strong-line parameters used to
derive metallicities are often calibrated employing sequences of models which do not follow
the distribution of metallicity obtained with more confident observational methods like,
for instance, the direct method (P\'erez-Montero \& D\'\i az, 2005), which is based
on the previous determination of the electron temperature of the ionized gas by means
of the fainter auroral emission lines.

HII galaxies usually represent the low-end of the mass-metallicity relation in different 
surveys.
In fact, the work of Lequeux et al. (1979), the first where the correlation between 
oxygen gas phase abundance and stellar mass was found,  is based on this
type of blue dwarf irregular galaxies.
Their spectra are characterized by very bright emission lines while the total luminosity 
and color are dominated by the presence of intense bursts of recent star formation 
(Sargent \& Searle, 1970; French, 1980). They are dwarf and compact so they seldom 
show chemical inhomogeneities. The metallicity distribution of HII galaxies
has an average value (12+log(O/H) $\sim$ 8.0, Terlevich et al., 1991)
sensibly lower than the rest of starburst galaxies and, in fact,
the objects with the lowest metal content in the Local Universe belong 
to the subclass of HII galaxies (IZw18, Searle \& Sargent, 1972).

Sloan Digital Sky Survey (SDSS) represents the largest sample of galaxies in the Local
Universe and it has been explored for the study of the mass-metallicity relation 
(Tremonti et al., 2004). Unfortunately, this survey presents some drawbacks that 
prevent a precise determination of the metallicity in the emission lines-like objects: 
1) Since the spectral coverage of the survey starts at 3800 {\AA}, the
sample objects at redshift $\sim$ 0 have not any detection of the [O{\sc ii}] $\lambda$ 3727 {\AA} 
line and hence no direct measurement of the total abundance of oxygen is possible. 
Only an estimate can be given in terms of the weaker auroral lines at $\lambda\lambda$ 
7319,7330 {\AA} but with a high degree of uncertainty (Kniazev et al., 2003).
2) Only a fraction of the objects show in their spectra the weak auroral emission 
lines necessary to estimate the electron temperature and, hence, to calculate chemical 
abundances following the direct method (Izotov et al., 2006).
3) The only auroral lines seen in the most part of this subsample is [O{\sc iii}] 
$\lambda$ 4363 {\AA} and, therefore, only the electron temperature of the high excitation 
zone can be derived. The total ionization structure of the nebula is then often 
obtained based on this temperature through a very simplified assumption 
(P\'erez-Montero \& D\'\i az, 2003). 
4) For the rest of the objects with no direct electron temperature derivation, 
a strong-line method is used, so the determination of metallicity is hugely sensitive
to the chosen calibration (Kewley \& Ellison, 2008). 5) The 3 arc secs aperture
of the SDSS fiber precludes a good determination of the physical properties of
the starburst region of these galaxies.

Double-arm spectrographs are suitable tools to observe the entire spectral range
between 3500 {\AA} and 1 micron in one single exposure at good spectral
resolution the star forming regions of
HII galaxies, avoiding the effects
due to a different covering of the object in different spectral ranges. 
In the case of compact HII galaxies, their bursts of star formation can be covered
entirely with a narrow slit. Assuming that different electron temperatures are measured
in the optical spectrum,
this implies that different excitation regions inside the ionized gas can be characterized,
leading to a more precise determination of the ionic chemical abundances.
There are some reported discrepancies between abundance determinations obtained from
bright collisional lines and the fainter recombination ones (Peimbert et al., 2007), which 
do not depend on the determination of the electron temperature. 
Nevertheless, the existence of these abundance discrepancy factors (ADFs) is still not conclusively established. 
In fact, in the case of HII galaxies, no significant deviations have been found between 
the electron temperatures derived from collisional lines and from 
the Balmer (H\"agele et al., 2006) or Paschen discontinuities (Guseva et al., 2006). 
Besides, the high ADFs derived from optical recombination lines are
sensibly lower than
those calculated from mid-infra-red lines 
(Wu et al., 2008; Lebouteiller et al., 2008) or from planetary nebulae
(Williams et al., 2008 ).


\begin{table*}
\centering
\caption[]{List of the modeled objects, with some additional basic information.}
\label{objects}
\begin{tabular} {c c c c c c l}
\hline
\hline
 \multicolumn{1}{c}{Object  ID}  &   hereafter ID &  Publication &  redshift & log L(H$\alpha$) &  12+log(O/H) & Other names\\
 &   & (Telescope)  &   &  (erg $\cdot$ s$^{-1}$)  \\
\hline
SDSS J002101.03+005248.1 & J0021   & H06 (WHT) &  0.098  & 42.13 &  8.10 $\pm$ 0.04 & UM228, SHOC 11 \\
SDSS J003218.60+150014.2 & J0032   & H06 (WHT) &  0.018  & 40.30 &  7.93 $\pm$ 0.03 & SHOC 22\\
SDSS J145506.06+380816.6 & J1455   & H08 (CAHA - 3.5 m.) &  0.028  & 40.84 & 7.94 $\pm$ 0.03 & CG 576 \\
SDSS J150909.03+454308.8 & J1509   & H08 (CAHA - 3.5 m.) &  0.048  & 41.26 & 8.19 $\pm$ 0.03 & CG 642 \\
SDSS J152817.18+395650.4 & J1528   & H08 (CAHA - 3.5 m.) &  0.064 & 41.63 & 8.17 $\pm$ 0.04 \\
SDSS J154054.31+565138.9 & J1540   & H08  (CAHA - 3.5 m.)&  0.011 & 39.65 & 8.07 $\pm$ 0.05 & SHOC 513  \\
SDSS J161623.53+470202.3 & J1616  & H08  (CAHA - 3.5 m.)&  0.002  & 38.93 & 8.01 $\pm$ 0.03 &\\
SDSS J162410.11-002202.5 & J1624  & H06  (WHT) &  0.031  & 41.48 & 8.05 $\pm$ 0.02 &SHOC 536 \\
SDSS J165712.75+321141.4 & J1657  & H08  (CAHA - 3.5 m.) & 0.038  & 40.74 & 7.99 $\pm$ 0.04 & \\
SDSS J172906.56+565319.4 & J1729  & H08  (CAHA - 3.5 m.) &  0.016  & 40.57 & 8.08 $\pm$ 0.04 & SHOC 575\\
\hline
\end{tabular}
\end{table*}

In this work we model the stellar content
 and the abundances of the most representative 
ions of a sample of HII galaxies previously observed in the SDSS and re-observed with
different double arm spectrographs in the spectral range between 3500 {\AA} and 1 micron
(H\"agele et al., 2006, hereafter H06 and H\"agele et al., 2008, hereafter H08).
The study of the stellar population in the star forming knots of these galaxies includes
underlying and ionizing populations. We show that the correct characterization of
the ionizing stellar population, including Wolf-Rayet stars,
and an appropriate geometry of the gas, dust content and the abundances of the main
elements are enough to reproduce with precision the intensities of the observed 
emission lines.
We demonstrate too that it is possible to 
draw a consistent picture of the thermal and ionization structure of the gas based on the measurement
of as many electron temperatures as possible and, hence, to calculate ionic abundances with 
high precision using optical to far red collisional emission lines.

In the next section we describe the sample of objects that we have modeled. 
In section 3 we describe our results concerning the stellar content through model fitting 
of the SED and the analysis of the Wolf-Rayet features. We also describe the 
photoionization models of the gas, once the underlying stellar population has 
been removed. In section 4, we discuss the star formation history and the 
physical properties obtained from the models and we compare them with observations.
In section 5, we summarize our results and present our conclusions. Finally, an appendix
has been added to compare the total abundances and ionization correction factors obtained
from models with those obtained in H06 and H08.


\section{The sample of studied objects}

The modeled galaxies were selected from the SDSS
using the implementation of the database in the INAOE Virtual Observatory
superserver\footnote{http://ov.inaoep.mx/} (see H06 and H08 for more details) .
We selected the brightest nearby
narrow emission line galaxies with very strong lines and large equivalent
widths of the H$\alpha$ line from the whole SDSS data release available when we
planned each observing run.  
Once AGN-like objects were removed  by BPT diagnostic diagrams (Baldwin, Phillips and Terlevich, 1981),
the sample was restricted to the largest H$\alpha$ flux and highest
signal-to-noise ratio objects (L\'opez, 2005).
The HII galaxies studied in this work comprise 10 objects belonging to
the final list and observed in two independent observing runs with different double-arm
spectrographs in order to improve the spectral range and the signal-to-noise.
Three out of them were observed with the ISIS spectrograph mounted on the
William Herschel Telescope (H06) and the
other 7 were observed with the TWIN instrument in the 3.5 m. CAHA telescope
(H08).  In Table \ref{objects} we list the galaxies,
with the names adopted in this work, the spectroscopic redshift
as listed in the SDSS catalogue and
their extinction-corrected H$\alpha$ luminosities in erg$\cdot$s$^{-1}$.
Distances have been derived from the measured redshifts for cosmological parameters: H$_{0}$ = 
73 km s$ ^{-1} $ Mpc$ ^{-1} $, $ \Omega_{matter} $ = 0.27, $ \Omega_{vaccum} $ = 0.73. 
We also list the oxygen abundance as calculated in H06 and H08
and, finally, other names by which the galaxies are also known.

Both observing runs had similar configurations. In WHT observations the
spectral range goes from 3200 {\AA} up to 10550 {\AA} with a spectral dispersion
of 2.5 {\AA}/px FWHM in the blue arm and 4.8 in the red. In CAHA observations, the spectral
range covers from 3400 up to 10400 {\AA}  (with a gap between 5700 and 5800 {\AA})
with spectral resolutions of 3.2 and 7.0 {\AA}/px FWHM in the blue and the red part respectively. 
The observations were taken at parallactic angle.
These configurations allow to detect simultaneously the bright emission lines
from [O{\sc ii}] at 3727 {\AA} up to [S{\sc iii}] at 9532 {\AA}, allowing the derivation 
of five different electron temperatures (t([O{\sc ii}]), t([O{\sc iii}]),
t([S{\sc ii}]),  t([S{\sc iii}]) and t([N{\sc ii}]) ) in three objects of the
sample and four temperatures (the same listed above 
with the exception of t([N{\sc ii}]) in the other seven).
The derivation of all these electron temperatures allows a more precise
determination of the thermal structure of the gas and, hence, of  
the ionic abundances of the detected species in the observed spectra.
Our sample galaxies have oxygen abundances in the range 7.93 $<$ 12+log(O/H) $<$ 8.19 and equivalent widths of H$\beta$ in
the range 83 {\AA} $<$ EW(H$\beta$) $<$ 171 {\AA}.

\section{Results}

\subsection{Model fitting of the stellar population}


\begin{figure*}
\begin{minipage}{170mm}
\centering{
\psfig{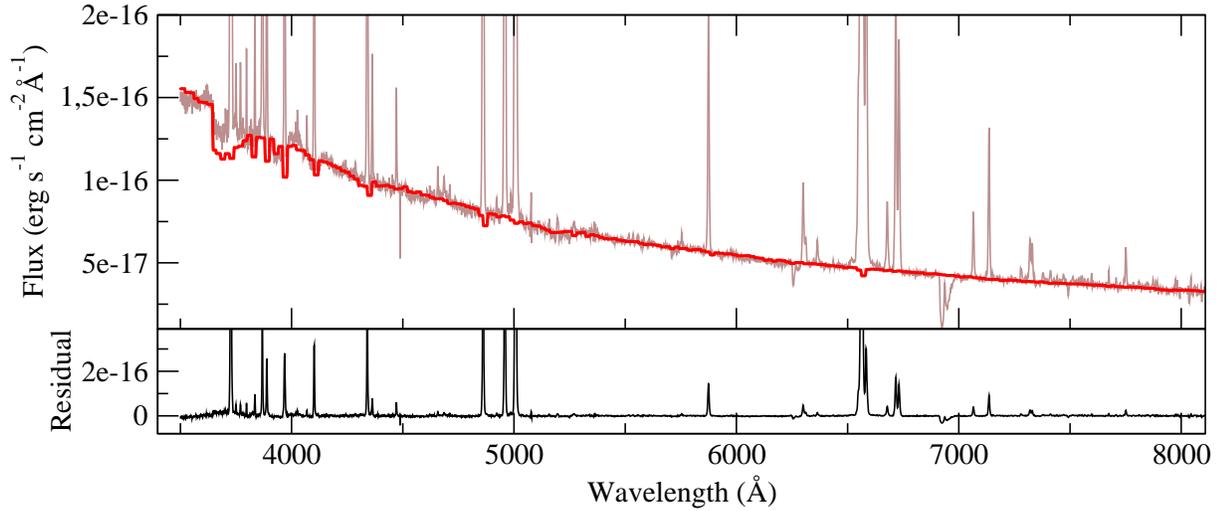}}
\caption{In the solid brown line, optical spectrum of the galaxy J0021 and comparison with the spectral fitting
carried out with the code STARLIGHT, using spectral libraries from Starburst99, represented by the solid red
line. In the lower panel, we show the residuals of the fit, which represents the nebular spectrum.}
\label{fit}
\end{minipage}
\end{figure*} 


\begin{figure}
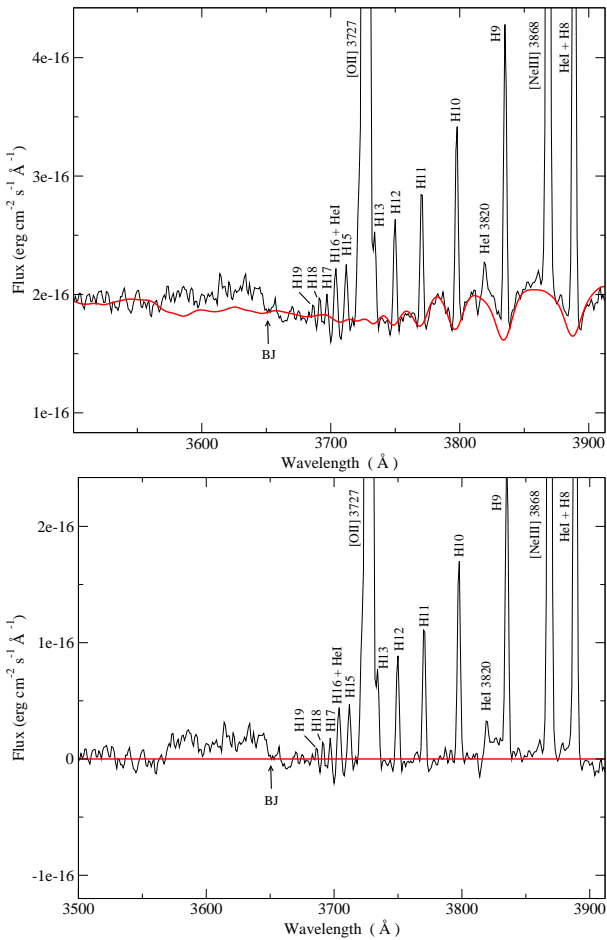

\begin{minipage}{85mm}
\centering{
\psfig{figure=mn-ghiimod.fg02u.eps,width=8cm,clip=}
\psfig{figure=mn-ghiimod.fg02d.eps,width=8cm,clip=}}

\caption{Upper panel, in black, optical spectrum of the galaxy J1624 in the spectral range 3500 - 3912 {\AA}, 
around the Balmer jump and the high order Balmer series, together with the fitting (red line) 
performed using the STARLIGHT spectral synthesis code. Lower panel, subtraction
of this fitting from the WHT spectrum in the same spectral range; the solid red line shows the zero level.}
\label{fit2}
\end{minipage}
\end{figure} 

Underlying stellar populations in starburst galaxies have several effects in their spectra that can affect
the measurement of emission lines. 
Firstly, since this population has a weight in the galaxy continuum, most of the times the
measurement of the equivalent widths of the emission lines is not a trustable estimator of the age of the ionizing stellar 
population (Terlevich et al., 2004). Secondly,
the presence of a conspicuous underlying stellar population depresses the Balmer 
and Paschen emission lines and do not allow to measure their 
fluxes with an acceptable accuracy (D\'\i az, 1988). 
This can alter those properties dependent on the fluxes of these recombination lines, like the reddening or the ionic abundances.
In the case of the helium recombination lines, this effect is present too, affecting the derivation
of the helium abundance (Olive \& Skillman, 2004).

We have subtracted from the observed spectra the spectral energy distribution of the underlying stellar
population found by the spectral synthesis code STARLIGHT
\footnote{The STARLIGHT project is supported by the Brazilian agencies CNPq, CAPES and 
FAPESP and by the France-Brazil CAPES/Cofecub program} (Cid Fernandes et al. 2004, 2005; Mateus
et al. 2005). STARLIGHT fits an observed continuum spectral energy distribution using a combination of 
multiple simple stellar population synthetic spectra (SSPs; also known as instantaneous burst) using a $\chi^2$ minimization procedure.
In order to keep the consistency with the photoionization models used to model the gas emission, 
we have chosen for our analysis the SSP spectra from the
Starburst99 libraries (Leitherer et al., 1999; V\'azquez \& Leitherer, 2005), based on stellar model atmospheres from
Smith et al. (2002), Geneva evolutionary tracks with high stellar mass loss (Meynet et al., 2004), 
a Kroupa Initial Mass Function (IMF; Kroupa, 2002) in two intervals (0.1-0.5 and 0.5-100 M$_\odot$)
with different exponents (1.3 and 2.3 respectively), the theoretical wind model (Leitherer et al., 1992)
 and a supernova cut-off of 8 M$_\odot$.
We have fixed the metallicity of the stellar populations to Z = 0.004 (= 1/5 Z$_\odot$) 
and Z = 0.008 (= 2/5 Z$_\odot$) depending on the
closest total oxygen abundance as measured using the direct method for each galaxy ( see H06 and H08).
The STARLIGHT code solves simultaneously the ages and relative contributions of the different SSPs and the average reddening. The reddening law from Cardelli, Clayton \& Matthews (1989)  has been used. Prior to the fitting procedure, the spectra were shifted to the rest-frame, and re-sampled to a wavelength interval of 1 {\AA} in the entire wavelength range between 3500 {\AA} and 9000 {\AA} by interpolation conserving flux, as required by the program. Bad pixels and emission lines were excluded from the final fits. 


\begin{figure*}
\begin{minipage}{170mm}
\centering{
\psfig{figure=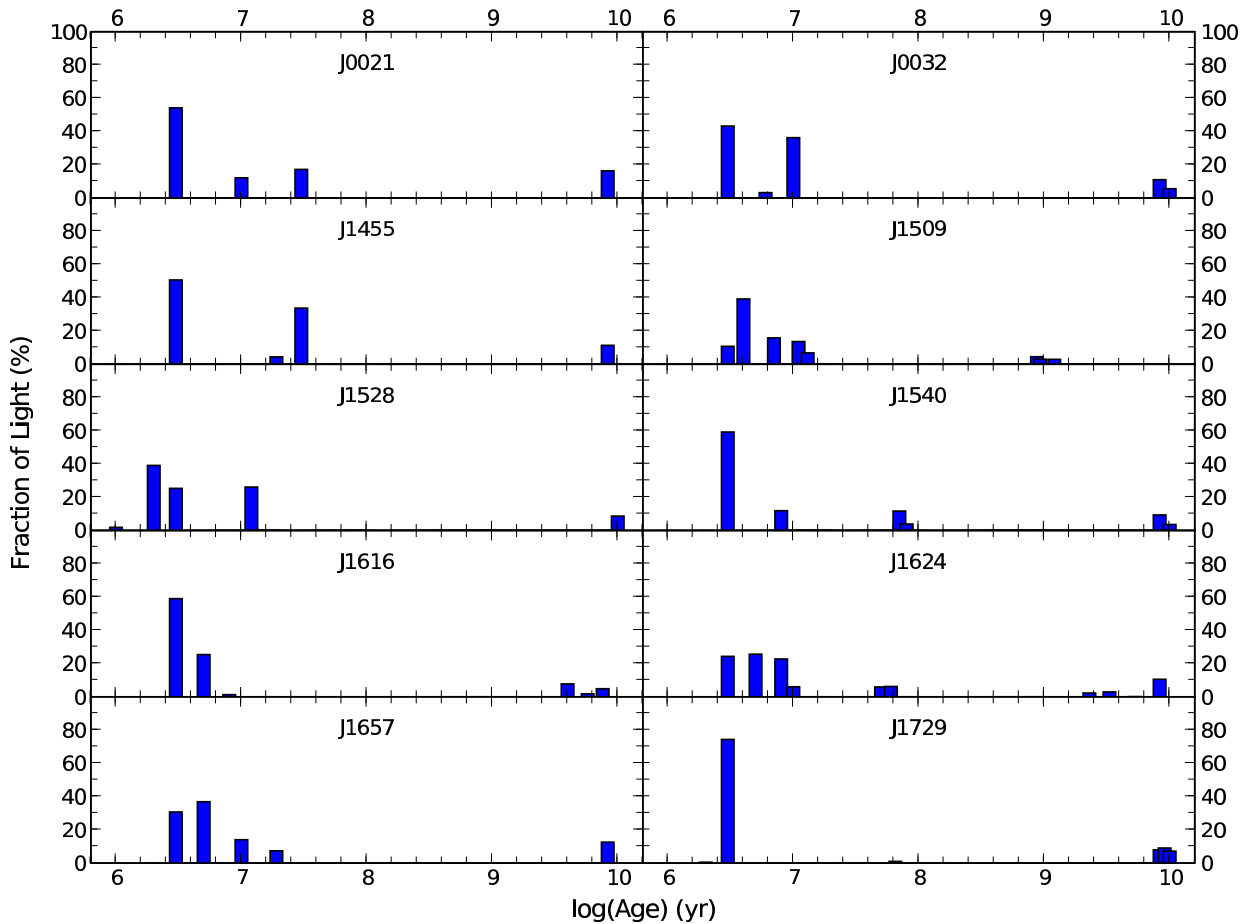,width=13cm,angle=-90,clip=}}
\caption{Histogram of the distribution of the percentage of visual light from each population in the best solution
found by STARLIGHT in the spectral fitting of the optical spectrum for each galaxy of the sample.}
\label{popgal}
\end{minipage}
\end{figure*} 

To illustrate the fitting results, we show in the upper panel of Figure \ref{fit} the spectrum of J0021 with the best fit found by STARLIGHT overimposed
and, in the lower panel of the same figure, the emission line spectrum once the underlying stellar continuum has been subtracted.
The measured intensities of the most prominent Balmer emission lines in this subtracted spectrum are, in all cases, equal within the errors to those reported in H06 and H08, where a pseudo-continuum was adopted.
It was also shown in those works that the measurement of the intensities of the
Balmer lines is consistent with a multi-gaussian fitting of the lines. 
This method fits independently the emission line and the absorption
feature, which is visible through their wings in some bright Balmer lines.
In the upper panel of Figure 2 we plot in detail the 
WHT spectrum of J1624 in the spectral range 3500-3912 {\AA}, around the Balmer
jump and the high order Balmer series, and in solid red
line we show the STARLIGHT fitting. 
In the lower panel of the same figure we show the subtraction of the fit from
the observed spectrum. We can appreciate
that small errors in the fit of the underlying stellar population could become
a great unquantifiable error in the emission line fluxes. 
Nevertheless, for the strongest emission lines, the
differences between the measurements done after the subtraction of the
STARLIGHT fit and those using the pseudo-continuum approximation are well below the
observational errors, giving almost the same result. It must be highlighted
that in this type of objects the absorption lines, used to fit the underlying
stellar populations, are mostly affected by the presence of emission
lines. There are a few that are not, such as some calcium and
magnesium absorption lines. To make the STARLIGHT fit it is necessary to mask
the contaminated lines, thus the fit is based on the continuum shape and takes
into account only a few absorption lines. Hence, it is not surprising that
the STARLIGHT results are not so good for this kind of objects.
It must be noted too that the STARLIGHT methodology is very advantageous for
statistical and comparative studies. Those studies only consider the results derived from
the strongest emission lines
which have small proportional errors. However, we must be careful when the aim
of our work is the detailed and precise study of a reduced sample of objects
because, as we have shown, we can not quantify the error introduced by the
fitting. Moreover, these errors are higher for the weakest lines, including
the auroral ones. Regarding helium absorption features the multigaussian approximation
is not possible in any line, because their wings are not visible. Nevertheless,
we have checked that the contribution of the depressed components of the emission lines
predicted by the STARLIGHT fit
are not larger than taking into account a slightly lower position of the continuum, as it is
explained in both H06 and H08.

>From this analysis we only take as a valuable input information for our tailor-made
photoionization models, the ratios predicted by STARLIGHT for the ionizing and
the underlying stellar populations in order to correct the equivalent width of H$\beta$ and
to take it as an estimate of the properties of the ionizing stellar population.


\begin{table*}
\centering
\caption[]{Properties of the stellar populations as obtained with STARLIGHT for each studied object. In the last two
columns we list the measured EW(H$\beta$) and its value corrected for contribution to the continuum from the young stellar component only.}
\label{fitting}
\begin{tabular} {c c c c c c c}
\hline
\hline
 \multicolumn{1}{c}{Object  ID}  &  A(V) &  log M$_*$ &  log M$_{ion *}$ & \% M$_{ion *}$ &  EW(H$\beta$) obs. &  EW(H$\beta$)$_{}cor$ \\
 & (mag)  & (M$_\odot$)  & (M$_\odot$)   &   & (\AA) & (\AA) \\
\hline

J0021   & 0.01 & 9.40 & 6.89 & 0.31 &  97 & 110 \\
J0032   & 0.17 & 8.15 & 5.53 & 0.24 &  90 & 119 \\
J1455   & 0.09 & 8.42 & 6.04 & 0.42 & 133 & 146 \\
J1509   & 0.58 & 8.08 & 6.96 & 7.57 & 123 & 135 \\
J1528   & 0.25 & 8.92 & 6.81 & 0.78 & 171 & 191 \\
J1540   & 0.42 & 7.51 & 5.22 & 0.52 & 122 & 124 \\
J1616  & -0.11 & 6.55 & 4.50 & 0.89 & 83 & 83 \\
J1624  & 0.60 & 8.88 & 5.73 & 0.07 & 101 & 107 \\
J1657  & 0.16 & 8.45 & 6.20 & 0.56 & 118 & 132 \\
J1729  & 0.27 & 8.65 & 6.07 & 0.26 & 126 & 126 \\

\hline
\end{tabular}
\end{table*}


Nevertheless, we can do a comparative analysis of the star formation histories of every galaxy as predicted by
the population synthesis fitting technique.
In Figure \ref{popgal}, we show the histogram of the age distribution of the 
contribution to the visual light of each stellar
population used by STARLIGHT to fit the optical SED for the ten studied galaxies 
As we can see, the blue light in these objects seems to be largely dominated by a very
young stellar population (younger than 10 Myr) which is responsible for the ionization of the gas.
At same time, there are different underlying stellar populations,  
with a component older
than 1 Gyr in all cases. This "old" component 
dominates the total stellar mass in all the objects, providing more than 99 \% 
of it, with the exception of J1509, for which this percentage is about 92 \%. At any rate, we find very similar
age distributions for the stellar populations in all the objects pointing to very similar star formation histories
with the presence of recursive starburst episodes resembling the one
observed at present.

In table \ref{fitting}, we summarize the properties of the best fitting to the optical spectrum of each galaxy as found by STARLIGHT. We list in columns 2 to 5 the visual extinction, A(V), the total stellar mass, M$_*$, the mass of the stellar population younger than 10 Myr, M$_{ion *}$, and  the percentage contribution of this to the total stellar mass, \% M$_{ion *}$. We also list in columns 6 and 7 the measured equivalent width of H$\beta$, EW(H$\beta$), and the corrected value, once the contribution to the continuum by the underlying stellar population has been subtracted.
As we can see, our procedure yields a negative value for the extinction in
J1616. Nevertheless, this is still consistent with the very low extinction found using
the Balmer decrements. 

We can check the accuracy of these fittings by comparing the stellar mass of the ionizing population as derived from
STARLIGHT, and the values of the same masses as derived from the extinction-corrected H$\alpha$ fluxes, using
the expression from D\'\i az (1998):

\begin{equation}
\log M_{ion*} = \log L(H\alpha) - 0.86 \cdot \log EW(H\beta) - 32.61
\end{equation}

\begin{figure}
\begin{minipage}{85mm}
\psfig{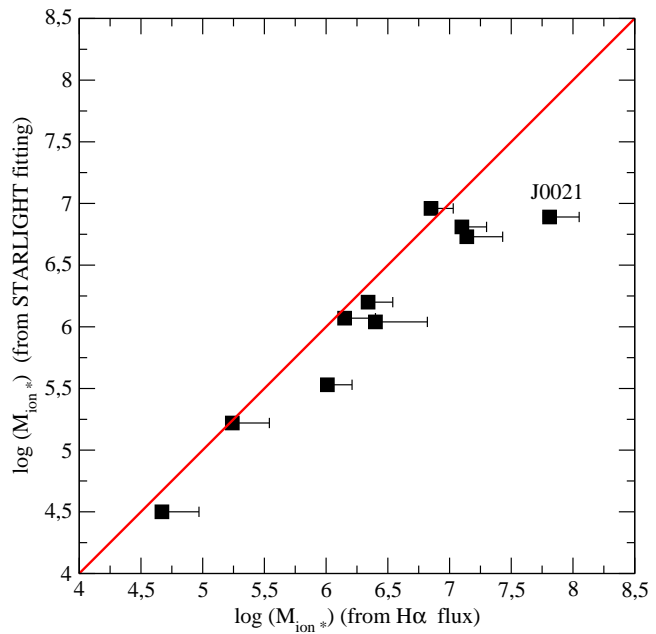}
\caption{Comparison between the ionizing stellar mass as derived from
the extinction-corrected H$\alpha$ fluxes and from SED fitting for
an age younger than 10 million years. The red solid line represents the 1:1 relation. Error bars to the right take into account the dust absorption factors obtained from the photoionization models described below. }
\label{mion}
\end{minipage}
\end{figure}

This comparison is shown in Figure \ref{mion}. It can be seen that the agreement between both derived values is very good for most of the studied objects. Only J0021 shows a higher young stellar mass when derived from its H$\alpha$ flux.
Horizontal error bars in the plot correspond to the underestimate of the H$\alpha$ flux due to the UV dust absorption. We have quantified this effect taking 
the absorption factors derived from the photoionization models described in section 3.3 below. 
Nevertheless, taking into account these corrected values of L(H$\alpha$) 
in the derivation of M$_{ion *}$ from (1) does not increase the ionizing stellar masses by more than 0.3 dex, except in the
case of J1455, with 0.42 dex.


\subsection{Detection and measurement of Wolf-Rayet bumps}


\begin{table*}
\begin{minipage}{180mm}
\centering
\caption[]{Observed extinction corrected H$\beta$ intensities and properties of the WR bumps for the objects of the sample. 
We list equivalent widths and  relative intensities, normalized to 100 times the H$\beta$ value, measured 
at 4650 {\AA} (blue bump, bb) and at 5808 {\AA} (red bump, rb). The corrected values take into account
the contribution of the underlying stellar population and the nebular emission to the continuum and
the dust absorption of the UV predicted by the photoionization models in each object.}
\label{wr}
\small
\begin{tabular} {c c c c c c c c c c}

\hline
\hline
 \multicolumn{1}{c}{Object  ID}  &  I(H$\beta$) & -EW(bb)  &  -EW(bb)$_{cor}$ & I(bb)\footnote{in units of 100$\cdot$I(H$\beta$)} & I(bb)$_{cor}^a$ & -EW(rb) & -EW(rb)$_{cor}$ & I(rb)$^a$ & I(rb)$_{cor}^a$ \\
 & (erg $\cdot$ s$^{-1}$ cm$^{-2}$) & (\AA)  &  (\AA) &    &    &  (\AA) & (\AA) &  \\
\hline

J0021   & 2.47 $\cdot$ 10$^{-14}$  & 3.9$\pm$0.9  &  {\em 5.8} &  1.4$\pm$0.3  &  {\em 0.8}  & --  & --  & -- & --  \\
J0032   & 1.23 $\cdot$ 10$^{-14}$   & 5.1$\pm$1.0  & {\em  6.7} &  4.9$\pm$1.0  & {\em 3.2}   & 1.3$\pm$0.6 & {\em 1.9} & 0.9$\pm$0.4   &  {\em 0.6}  \\
J1455   & 1.49 $\cdot$ 10$^{-14}$  & --           & --           & --            & --   & -- & --  & --  & --       \\
J1509   & 1.35 $\cdot$ 10$^{-14}$  & 9.1$\pm$2.0 & {\em 12.0} & 8.3$\pm$1.8  & {\em 5.3} & 1.6$\pm$0.5 & {\em 2.2} & 1.1$\pm$0.4   & {\em 0.7}   \\
J1528   & 1.73 $\cdot$ 10$^{-14}$  & --           & --           & --            & --       & --  & --   & --  & -- \\
J1540   & 5.30 $\cdot$ 10$^{-15}$  & 10.3$\pm$2.8 & {\em 16.5} & 9.8$\pm$2.6  & {\em 9.5} & -- & -- & -- & --     \\
J1616   & 1.36 $\cdot$ 10$^{-14}$  & 5.5$\pm$3.8 &  {\em 6.6} &  6.9$\pm$4.8  & {\em 3.3}  & --            & --   & --  & --           \\
J1624   & 5.03 $\cdot$ 10$^{-14}$  & 7.4$\pm$1.5 &  {\em 13.1} &  2.6$\pm$0.6  &  {\em 1.3}  & 2.8$\pm$0.9 & {\em  3.8} & 0.8$\pm$0.3   &  {\em  0.4}  \\
J1657   & 6.30 $\cdot$ 10$^{-15}$  & --           & --           & --            & --   & --  & --  & -- & --           \\
J1729   & 2.40 $\cdot$ 10$^{-14}$  & 6.5$\pm$2.1 & {\em 9.4} &  6.5$\pm$2.1  & {\em 3.4}  & --            & --  & --  & --           \\
\hline
\end{tabular}
\end{minipage}
\end{table*}


Wolf-Rayet (WR) stars appear in the first stages after the main sequence phase of the evolution of massive stars, so they
are already visible soon after the beggining of a burst of star formation (approximately 2 Myr) 
and for a relatively short period of time (about 3 Myr, on average). The existence
of this WR stellar population can be very useful for the study and characterization of the ionizing stellar
population in starburst galaxies ({\em e.g.} P\'erez-Montero \& D\'\i az, 2007).
The strength of the stellar winds produced by these stars can be sometimes measured in the integrated spectra of the
starburst galaxies, which are then identified as WR galaxies (Conti, 1991). The presence of the emission broad features ("bumps") produced by WR stars are therefore associated to the existence of processes of intense star formation. 
These WR features are the blue bump, centered
at 4650 {\AA} and produced mainly by broad emission lines of  N{\sc v} at 4605, 4620 {\AA}, N{\sc iii}
4634, 4640 {\AA}, C{\sc iii/iv} 4650, 4658 {\AA} and He{\sc ii} at 4686 {\AA}, and 
the red bump, usually fainter, centered at 5808 {\AA} and emitted mainly by C{\sc iii}.

We have detected the blue bump in seven of the ten galaxies of our sample, while the red bump
has been detected in only three of these seven galaxies. All the three galaxies observed with WHT
had already been identified as WR galaxies in H06 and have also been included 
in the catalogue of WR galaxies of the SDSS (Brinchmann et al., 2008). 
The other four galaxies with a detection of WR features, observed with the 3.5 CAHA telescope are not listed
in that catalogue. An example of the WR emission can be seen in Figure 6 of H06 for the J1624 galaxy.

We have measured the bumps in the following way. First we have adopted a continuum shape and we have
measured the broad emission over this continuum at the wavelengths of the blue bump, at 4650 {\AA}, and
the red bump, at 5808 {\AA}. Finally, we have subtracted the emission of the narrower lines not emitted by
the WR stars winds, which are [Fe{\sc iii}] 4658 {\AA}, the narrow component of He{\sc ii} 4686 {\AA}, 
He{\sc i} + [Ar{\sc iv}] 4711 {\AA} and [Ar{\sc iv}] 4740 {\AA} in the blue bump.
The equivalent widths and the deredenned relative intensities of the blue and red bumps
are shown in Table \ref{wr}.  Both have been corrected in order to study the properties
of the ionizing stellar population. In the case of EWs, we have subtracted the contribution
of the underlying stellar population as derived from STARLIGHT and from the nebular
continuum as derived from the photoionization models described below. In the case of
the relative intensities, we have taken into account the dust absorbed component of
H$\beta$ derived from the same photoionization models for each object.



\subsection{Photoionization models of the ionized gas}

Photoionization models have been calculated using the code CLOUDY v. 06.02c (Ferland et al., 1998)
with the same atomic coefficients used to derive ionic abundances in H06 and H08 and the
recent dielectronic recombination rate coefficients from Badnell (2006).
We have taken the same Starburst99 stellar libraries used by STARLIGHT to fit the observed SEDs.
We have assumed different initial input parameters and then, we have followed an iterative
method to fit the relative intensities of the strongest emission lines in the integrated spectra,
along with other observable properties. 

We have assumed a constant star formation history, with a star formation rate calculated according
to the measured and corrected number of ionizing photons. The age of this burst of star formation
has been limited to the range 1-10 Myr in order to reproduce the corrected equivalent width of H$\beta$ once the
older stellar populations have been removed, except in the case of J1540 for which a burst duration of 15.85 Myr 
has been considered in order to reproduce the observed EW(H$\beta$).
Total abundances have been set to match the values derived in H06 and H08, including 
He, O, N, S, Ne, Ar and Fe. The abundances of the rest
of the elements have been scaled to the measured oxygen abundance of each galaxy and taking as
reference the photosphere solar values given by Asplund et al. (2005).
The inner radii of the ionized regions have been chosen to reproduce a thick shell in all cases. This kind of
geometry reproduces best the ionization structure of oxygen and sulphur simultaneously
 (P\'erez-Montero \& D\'\i az, 2007).
We have assumed a constant density equal to that derived from the measured [S{\sc ii}] emission line ratio
($\sim$ 100 particles per cm$^{3}$ in all the objects).
We have kept as free parameters the filling factor and the amount of dust. 
The inclusion of dust allows to fit correctly the measured electron temperatures because it is a fundamental 
component to reproduce the thermal balance in the gas.
The heating of the dust can affect the equilibrium between
the cooling and heating of the gas and, thus,  the electron temperature inside the nebula.
We have assumed the default grain properties given by Cloudy 06, which has essentially the properties 
of the interstellar medium and follows a MRN (Mathis, Rumpl \& Norsieck 1997) grains size distribution. 

After some first estimative guesses
of the UV absorption factor due to the internal extinction, $f_d$, we have recalculated the age of the cluster and the number
of ionizing photons to match the observed values. Finally, we have adjusted the chemical abundances to fit
the relative intensities of the emission lines. 

The average number of models needed to fit the observable properties of the galaxies with a reasonable
accuracy is about 50 for each galaxy. In Table \ref{lineas} we show the intensities of the most representative emission lines as measured in
H06 and H08 compared to the values obtained with the best model for each object.
The agreement in most cases is excellent, even for the faint auroral lines.


\begin{table*}
\caption{Comparison between the observed intensities of the most representative emission
lines in terms of 100 $\cdot$ F(H$\beta)$ and the
reproduced intensities by the best model obtained for each object.}
\label{lineas}
\begin{center}
\scriptsize
\begin{tabular}{lcccccccccc}
\hline
\multicolumn{1}{c}{$\lambda$  ({\AA})}  & \multicolumn{2}{c}{J0021} &
\multicolumn{2}{c}{J0032}  & \multicolumn{2}{c}{J1455} &
\multicolumn{2}{c}{J1509} & \multicolumn{2}{c}{J1528} \\
 & Observed  & {\em Model} &  Observed  & {\em Model} & Observed  & {\em Model} & Observed  & 
{\em Model} & Observed  &  {\em Model} \\
\hline
3727  [O{\sc ii}]             & 163.4 $\pm$ 1.9 &   {\em 162.9 }   & 157.3 $\pm$ 1.4 &    {\em 157.9  }   & 111.5 $\pm$ 1.6 &    {\em 117.5 }  & 153.2 $\pm$ 1.8 &    {\em 163.6 }  & 228.8 $\pm$ 2.9  &    {\em 230.5}  \\
3868  [Ne{\sc iii}]           &  38.8 $\pm$ 0.8 &    {\em 40.6}    &  37.2 $\pm$ 1.0 &     {\em 39.2  }   &  47.9 $\pm$ 1.1 &     {\em 50.1}   &  35.0 $\pm$ 1.1 &    {\em 34.1}   &   47.2 $\pm$ 1.8  &   {\em   46.6 } \\
4363  [O{\sc iii}]            &   5.6 $\pm$ 0.3 &     {\em 5.7 }   &   6.2 $\pm$ 0.3 &      {\em 6.0  }   &  10.2 $\pm$ 0.4 &     {\em 10.3}   &    4.2 $\pm$ 0.2 &    {\em  4.1}   &    5.0 $\pm$ 0.2  &       {\em 5.0}  \\
4658  [Fe{\sc iii}]           &   1.0 $\pm$ 0.1 &     {\em 1.1}    &   0.9 $\pm$ 0.1 &      {\em 0.9  }   &   --  &     {\em  1.0}   &   1.1 $\pm$ 0.1 &    {\em 0.9 }   &   1.3 $\pm$ 0.2  &    {\em   1.4}   \\ 
4959  [O{\sc iii}]            & 153.2 $\pm$ 1.3 &   {\em 140.9 }   & 158.2 $\pm$ 1.3 &    {\em 156.1 }    & 204.6 $\pm$ 1.3 &    {\em 192.5 }   & 167.5 $\pm$ 1.5 &    {\em 154.0 }   & 165.6 $\pm$ 1.7  &   {\em  162.2}   \\
5007  [O{\sc iii}]            & 433.4 $\pm$ 2.6 &   {\em 424.0 }   & 460.7 $\pm$ 2.4 &    {\em 469.9  }   & 613.6 $\pm$ 3.4 &  {\em   579.4 }   & 499.4 $\pm$ 1.5 &   {\em 463.4 }   & 489.3 $\pm$ 2.92  &   {\em  488.1}   \\
5876  He{\sc i}               &  12.7 $\pm$ 0.7 &    {\em 13.1}    &  12.4 $\pm$ 0.6 &    {\em  11.4 }    &   11.4 $\pm$ 0.35 &    {\em  10.2}    &  12.68 $\pm$ 0.59 &   {\em  12.6 }   &  12.27 $\pm$ 0.36  &     {\em 11.8 }   \\
6312  [S{\sc iii}]            &   1.2 $\pm$ 0.1 &     {\em 1.6}    &   2.3 $\pm$ 0.1 &      {\em 2.1 }    &    1.6 $\pm$ 0.1 &    {\em   1.8 }  &   1.6 $\pm$ 0.1 &    {\em  2.0 }  &    1.7 $\pm$ 0.1  &       {\em 1.6 } \\
6548  [N {\sc ii}]             &     ---         &     {\em 9.5 }   &   3.7 $\pm$ 0.2 &     {\em  3.6  }   &   2.7 $\pm$ 0.2 &     {\em  3.0 }  &   5.4 $\pm$ 0.3 &    {\em  5.0  } &   7.2 $\pm$ 0.5  &       {\em 7.0 } \\
6584  [N{\sc ii}]             &  26.0 $\pm$ 0.9 &    {\em 28.1}    &  11.2 $\pm$ 0.5 &    {\em  11.2 }    &   7.9 $\pm$ 0.2 &     {\em  8.8 }  &  13.9 $\pm$ 0.4 &   {\em  14.6}   &  19.6 $\pm$ 0.4  &     {\em 20.6 } \\
6717  [S{\sc ii} ]             &  13.6 $\pm$ 0.5 &    {\em 14.5}    & 17.3 $\pm$ 0.5 &    {\em  16.9  }   &   10.0 $\pm$ 0.3 &     {\em  9.6 }    &   19.7 $\pm$ 2.1 &    {\em 18.3 }  &  19.2 $\pm$ 0.8  &   {\em   18.4 }    \\
6731  [S{\sc ii}]             &  10.7 $\pm$ 0.5 &    {\em 11.0 }   &  12.5 $\pm$ 0.3 &    {\em  12.7  }   &   7.9 $\pm$ 0.2 &    {\em   7.2 } &   14.9 $\pm$ 0.4 &    {\em 13.9 } &  14.2 $\pm$ 0.6  &      {\em 14.0 } \\
7136  [Ar{\sc iii}]           &   6.2 $\pm$ 0.2 &     {\em 6.5 }   &   9.2 $\pm$ 0.3 &      {\em 8.2 }    &    6.62 $\pm$ 0.3 &    {\em   7.2}  &   9.82 $\pm$ 0.29 &    {\em  9.8 } &       -----        &       {\em 8.9} \\
7319  [O{\sc ii}]             &   2.0 $\pm$ 0.1 &     {\em 2.4 }   &   2.5 $\pm$ 0.1 &      {\em 2.4 }    &   1.8 $\pm$ 0.1 &     {\em  1.9}  &   2.0 $\pm$ 0.1 &    {\em  2.2 } &   3.0 $\pm$ 0.2  &       {\em 3.1} \\
7330  [O{\sc ii}]             &   1.3 $\pm$ 0.1 &     {\em 2.0 }   &   1.9 $\pm$ 0.1 &      {\em 1.9  }   &   1.4 $\pm$ 0.1 &      {\em 1.5 } &   1.7 $\pm$ 0.1 &   {\em   1.8 } &   2.4 $\pm$ 0.1  &       {\em 2.6} \\
9069  [S{\sc iii}]            &  11.9 $\pm$ 0.6 &    {\em 15.0  }  &  21.7 $\pm$ 1.0 &     {\em 19.9 }      &  11.5 $\pm$ 0.6 &    {\em  14.1 } &   25.5 $\pm$ 1.2 &   {\em  23.9}   &  16.9 $\pm$ 1.3  &    {\em  17.3 } \\
9532  [S{\sc iii}]            &  23.9 $\pm$ 0.9 &    {\em 37.2  }  &  44.5 $\pm$ 3.8 &     {\em 47.8 }    &  32.8 $\pm$ 1.5 &    {\em  34.9}   &  51.8 $\pm$ 2.6 &    {\em 59.3 }  &  40.5 $\pm$ 2.9  &     {\em 43.0 } \\
\hline
\hline
\multicolumn{1}{c}{$\lambda$  ({\AA})}  & \multicolumn{2}{c}{J1540} &
\multicolumn{2}{c}{J1616}  & \multicolumn{2}{c}{J1624} &
\multicolumn{2}{c}{J1657} & \multicolumn{2}{c}{J1729} \\
 & Observed  &  {\em Model} &  Observed  & {\em Model} & Observed  & {\em Model} & Observed  & 
{\em Model }& Observed  & {\em Model} \\
\hline
 3727 [O{\sc ii}]             & 217.9 $\pm$ 2.6 & {\em  215.3}  & 84.9 $\pm$ 2.0 & {\em  84.1 }  & 147.1 $\pm$ 2.2 &    {\em 157.6}  & 188.3 $\pm$ 2.3  & {\em  203.2 }   & 176.2 $\pm$ 2.4 &     {\em 164.4 }  \\
 3868 [Ne{\sc iii}]           &  21.4 $\pm$ 0.8 &   {\em 21.8}  & 41.1 $\pm$ 1.7 &  {\em 44.0}   &  42.8 $\pm$ 1.1 &     {\em 41.4}  & 32.6 $\pm$ 1.3  &  {\em  33.6 }   &  47.9 $\pm$ 1.7 &      {\em 50.0 }  \\
 4363 [O{\sc iii}]            &   2.9 $\pm$ 0.2 &    {\em 2.9 } &  8.5 $\pm$ 0.3 &   {\em 9.1 }  &   7.0 $\pm$ 0.2 &      {\em 7.1}  &  5.2 $\pm$ 0.2  &  {\em   5.2}    &     6.6 $\pm$ 0.3 &    {\em    6.5 }  \\
 4658 [Fe{\sc iii}]           &   0.8 $\pm$ 0.1 &    {\em 0.8}  &    -- &   {\em 6.9 }           &   0.9 $\pm$ 0.1  &     {\em  0.7 } &  1.1 $\pm$ 0.2  &  {\em   1.1 }   &    0.9 $\pm$ 0.1 &    {\em    0.9 }  \\
 4959 [O{\sc iii}]            &  104.8 $\pm$ 0.8 &  {\em 115.3}  & 204.9 $\pm$ 1.5 & {\em 204.8 }  & 193.8 $\pm$ 1.4 &   {\em  177.5}  & 143.3 $\pm$ 1.3 & {\em  130.0}    & 171.0 $\pm$ 1.5 &    {\em  184.3  } \\
 5007 [O{\sc iii}]            & 309.4 $\pm$ 1.9 &  {\em 347.2 } & 615.2 $\pm$ 3.7 & {\em 616.5}   & 564.2 $\pm$ 2.8 &   {\em  534.4}  & 430.8 $\pm$ 2.4  & {\em  391.4 }   & 515.4 $\pm$ 4.2 &  {\em    554.7 }  \\
 5876 He{\sc i}               &  11.6 $\pm$ 0.4 &   {\em 12.8}  & 10.6 $\pm$ 0.8 &  {\em 11.2 }  &  13.5 $\pm$ 0.3 &     {\em 14.0}  & 11.2 $\pm$ 0.4  &  {\em  11.4 }   &  12.6 $\pm$ 0.4 &      {\em 13.1  } \\
 6312 [S{\sc iii}]            &    1.4 $\pm$ 0.1 &    {\em 1.8 } &  1.9 $\pm$ 0.1 &   {\em 1.9 }  &   1.8 $\pm$ 0.1 &      {\em 1.7}  &  2.0 $\pm$ 0.1  &  {\em   1.8 }   &   1.8 $\pm$ 0.1 &     {\em   1.8  } \\
 6548 [N{\sc ii}]             &   7.4 $\pm$ 0.3 &    {\em 5.5 } &   1.6 $\pm$ 0.1 &   {\em 1.3 }  &   3.4 $\pm$ 0.2 &      {\em 3.0 } &  4.6 $\pm$ 0.2  &  {\em   4.9}    &   7.7 $\pm$ 0.2 &    {\em    7.6}   \\
 6584 [N{\sc ii}]             &  21.2 $\pm$ 0.7 &   {\em 16.4}  &  4.3 $\pm$ 0.2 &   {\em 4.0 }  &   9.3 $\pm$ 0.2 &      {\em 8.9 }    & 14.3 $\pm$ 0.5  &  {\em  14.6 }   &  21.9 $\pm$ 0.5 &      {\em 22.6 }  \\
 6717 [S{\sc ii}]             &  26.1 $\pm$ 0.5 &   {\em 25.2}  &  7.7 $\pm$ 0.2 &   {\em 7.9}   &  13.6 $\pm$ 0.3 &     {\em 12.6 }   & 22.1 $\pm$ 0.6  &  {\em  19.6 }   &  12.9 $\pm$ 0.3 &      14.4  \\
 6731 [S{\sc ii}]             &   19.2 $\pm$ 0.5 &   {\em 19.1}  &  5.8 $\pm$ 0.2 &   {\em 6.0 }  &   9.9 $\pm$ 0.3 &      {\em 9.5 }   & 16.0 $\pm$ 0.4  &  {\em  14.8 }   &   10.1 $\pm$ 0.3 &      {\em 10.9 }  \\
 7136 [Ar{\sc iii}]           &   8.9 $\pm$ 0.5 &    {\em 8.8 } &  7.4 $\pm$ 0.4 &   {\em 7.1 }  &   8.0 $\pm$ 0.3 &      {\em 8.0 }   &  7.2 $\pm$ 0.3  &  {\em   6.3 }   &   8.6 $\pm$ 0.4 &     {\em   9.5  }    \\
 7319 [O{\sc ii}]             &   2.7 $\pm$ 0.2 &    {\em 2.9}  &  1.4 $\pm$ 0.1 &   {\em 1.3 }  &   2.1 $\pm$ 0.1 &     {\em  2.4 }   &  3.0 $\pm$ 0.2  & {\em    3.0}    &   2.3 $\pm$ 0.1 &       2.4  \\
 7330 [O{\sc ii}]             &   2.2 $\pm$ 0.1 &    {\em 2.3}  &   0.9 $\pm$ 0.1 &   {\em 1.1}   &   1.8 $\pm$ 0.1 &     {\em  1.9  }  &  2.1 $\pm$ 0.1  &  {\em   2.4}    &   1.9 $\pm$ 0.1 &       1.9  \\
 9069 [S{\sc iii}]            &  21.0 $\pm$ 0.6 &   {\em 22.0}  & 16.5 $\pm$ 0.6 &  {\em 15.9 }  &   7.0 $\pm$ 0.5 &     {\em 15.4  }  &  14.0 $\pm$ 1.0  &  {\em  16.6 }   &  20.9 $\pm$ 1.0 &      {\em 18.1   } \\
 9532 [S{\sc iii}]            &  53.3 $\pm$ 3.3 &   {\em 54.7 } & 40.1 $\pm$ 1.5 &  {\em 39.5}   &  40.1 $\pm$ 1.9 &     {\em 38.2  }  & 36.7 $\pm$ 2.6  &  {\em  41.2}    &   47.2 $\pm$ 2.5 &      {\em 45.0  }  \\
\hline

\end{tabular}
\end{center}
\end{table*}


In Table \ref{abs}, we show the different line temperatures obtained from the photoionization models 
compared to those derived in H06 and H08. In the same table,  we show
the elemental ionic abundances derived using the corresponding electron temperature representative of the region where the ion is formed: t([O{\sc iii}]) for O$^{2+}$, Ne$^{2+}$ and Fe$^{2+}$; t([S{\sc iii}]) for S$^{2+}$ and Ar$^{2+}$; t([O{\sc ii}]) for O$^{+}$ and t([N{\sc ii}]) for N$^{+}$ when t([N{\sc ii}]) is
available or t([O{\sc ii}]) in the other cases. 
These abundances are compared with the ionic abundances obtained from the corresponding models.
We also show the temperature from the oxygen recombination lines predicted by our models in the
high excitation region and the temperature fluctuations ($t^2$, Peimbert, 1967) obtained from
the difference between the temperatures derived from recombination and collisional emission lines.
It can be seen that the predicted fluctuations are practically zero and,
therefore, no ADFs are expected according to our models. 
The comparison between the model predictions and the temperature fluctuations inferred
from the observed values of t([O{\sc iii}]) and  the Balmer jump temperature
for the three objects observed with the WHT gives
compatible results, with the exception of J0032, whose $t^2$ is higher as obtained from the observations.


\begin{table*}
\begin{minipage}{180mm}
\begin{center}
\scriptsize
\caption[]{Comparison between the derived electron temperatures and ionic abundances
derived in H06 and H08, and the values obtained from the best tailor-made photoionization  model 
for each object.}
\label{abs}
\begin{tabular} {l c c c c c c c c c c}
\hline
\hline
 \multicolumn{1}{c}{}  & \multicolumn{2}{c}{J0021} & \multicolumn{2}{c}{J0032} & \multicolumn{2}{c}{J1455} &
\multicolumn{2}{c}{J1509} &\multicolumn{2}{c}{J1528} \\
    &  Derived  &  {\em Model } &  Derived  & {\em  Model }   &  Derived  & {\em  Model }  &  Derived  & {\em  Model }   &  Derived  &  Model \\
\hline

T([O{\sc ii}]) ($K$)  &  10300 $\pm$ 200 & {\em 12303 } & 13500 $\pm$ 400 & {\em 12388}  & 13300 $\pm$ 700 &{\em  13062}  & 11800 $\pm$ 500 &{\em  11561 } & 11700 $\pm$ 500 &{\em  11695 }  \\
12+log(O$^+$/H$^+$) & 7.73 $\pm$ 0.06 & {\em 7.44}   & 7.15 $\pm$ 0.06 & {\em 7.43}   & 7.16 $\pm$ 0.09 & {\em 7.20}   & 7.48 $\pm$ 0.08 &{\em  7.57}   & 7.67 $\pm$ 0.09 & {\em 7.68} \\
\\
T([O{\sc iii}]) ($K$)  &  12500 $\pm$ 200 & {\em 12882 } & 12800 $\pm$ 200 & {\em 12531}  & 14000 $\pm$ 200 & {\em 14421}  & 10900 $\pm$ 100 & {\em 11066}  & 11600 $\pm$ 100 & {\em 11695} \\
T$_r$([O{\sc iii}]) ($K$)  & -- & {\em 13000 } & -- & {\em 12900}  & -- & {\em 14700}  & -- & {\em 11200}  & -- &{\em  11800}  \\
$t^2$ & 0.004$^{+0.044}_{-0.004}$ &{\em  0.002}  & 0.066$\pm$0.026 & {\em 0.001}  & -- & {\em 0.005}  & -- & {\em 0.002}  & -- & {\em 0.001}  \\
12+log(O$^{2+}$/H$^+$) & 7.86 $\pm$ 0.02 & {\em 7.81}  & 7.86 $\pm$ 0.02 & {\em 7.88}   & 7.87 $\pm$ 0.02 & {\em 7.83}   & 8.10 $\pm$ 0.02 & {\em 8.03}   & 8.00 $\pm$ 0.02 & {\em 7.98}  \\
12+log(Ne$^{2+}$/H$^+$) & 7.27 $\pm$ 0.03 &{\em  7.16}  & 7.30 $\pm$ 0.06 & {\em 7.18}   & 7.20 $\pm$ 0.03 & {\em 7.12}   & 7.44 $\pm$ 0.03 & {\em 7.28}   & 7.47 $\pm$ 0.04 &{\em  7.34}  \\
12+log(Fe$^{2+}$/H$^+$) & 5.50 $\pm$ 0.05 & {\em 5.39}  & 5.42 $\pm$ 0.06 & {\em 5.33}   & 4.81 $\pm$ 0.08 & {\em 5.27}   & 5.71 $\pm$ 0.06 & {\em 5.45}   & 5.69 $\pm$ 0.07 & {\em 5.58} \\
\\
T([N{\sc ii}]) ($K$)  &  11900 $\pm$ 500 & {\em 12078} & -- & {\em 12162} & -- & {\em 12647} & -- & {\em 11482} & -- & {\em 11561}  \\
12+log(N$^+$/H$^+$) & 6.68 $\pm$ 0.04 & {\em 6.50} &  6.03 & {\em 6.09}  & 5.90 $\pm$ 0.06 & {\em 5.94} & 6.28 $\pm$ 0.06 & {\em 6.30}  & 6.43 $\pm$ 0.06 & {\em 6.42}\\
\\
T([S{\sc ii}]) ($K$)  &  8600 $\pm$ 600 & {\em 11776}  & 10300 $\pm$ 500 & {\em 11885}  & 13100 $\pm$ 1100 & {\em 12388}  & 8900 $\pm$ 700 & {\em 11220}  & 9900 $\pm$ 700 & {\em 11246} \\
12+log(S$^+$/H$^+$) & 5.93 $\pm$ 0.11 & {\em 5.56}  & 5.80 $\pm$ 0.06 & {\em 5.63}   & 5.36 $\pm$ 0.06 &{\em  5.32}  & 6.02 $\pm$ 0.10 & {\em 5.74}   & 5.89 $\pm$ 0.10 & {\em 5.71} \\
\\
T([S{\sc iii}]) ($K$) &  13100 $\pm$ 500 & {\em 12706}  & 13600 $\pm$ 700 & {\em 12556}  & 13700 $\pm$ 500 & {\em 14060}  & 10200 $\pm$ 400 & {\em 11194}  & 12100 $\pm$ 600 & {\em 11668} \\
12+log(S$^{2+}$/H$^+$) & 5.91 $\pm$ 0.05 & {\em 6.14}  & 6.16 $\pm$ 0.06 & {\em 6.26}   & 5.98 $\pm$ 0.05 & {\em 6.05}   & 6.44 $\pm$ 0.05 & {\em 6.44}   & 6.18 $\pm$ 0.07 & {\em 6.26} \\
12+log(Ar$^{2+}$/H$^+$) & 5.50 $\pm$ 0.05 &{\em  5.52}  & 5.65 $\pm$ 0.06 & {\em 5.62}   & 5.50 $\pm$ 0.05 & {\em 5.49}   & 5.94 $\pm$ 0.05 & {\em 5.81}   & 5.73 $\pm$ 0.09 & {\em 5.72} \\
\\
log(He$^+$/H$^+$) & -1.05 $\pm$ 0.03 & {\em -0.99 } &  -1.06 $ \pm$ 0.04 & {\em -1.07}  & -1.05 $\pm$ 0.05 & {\em -1.11}   & -1.00 $\pm$ 0.05 & {\em -1.01}  & -1.02 $\pm$ 0.05 & {\em -1.03}  \\
\hline
 \multicolumn{1}{c}{}  & \multicolumn{2}{c}{J1540} & \multicolumn{2}{c}{J1616} & \multicolumn{2}{c}{J1624} &
\multicolumn{2}{c}{J1657} &\multicolumn{2}{c}{J1729} \\
    &  Derived  &  {\em Model} &  Derived  &  {\em Model } &  Derived  &  {\em Model } &  Derived  & {\em  Model }&  Derived  & {\em  Model} \\
\hline

T([O{\sc ii}]) ($K$)  &  11500 $\pm$ 600 & {\em 11402 }& 12900 $\pm$ 900 &{\em  12853} & 13100 $\pm$ 300 &{\em  12369} & 13300 $\pm$ 700 & {\em 12246} & 11600 $\pm$ 400 & {\em 12162} \\
12+log(O$^+$/H$^+$) & 7.67 $\pm$ 0.09 & {\em 7.71}  & 7.07 $\pm$ 0.12 & {\em 7.09}  & 7.18 $\pm$ 0.06 & {\em 7.42}  & 7.37 $\pm$ 0.09 & {\em 7.54}  & 7.57 $\pm$ 0.07 & {\em 7.48}\\
\\
T([O{\sc iii}]) ($K$)  &  11300 $\pm$ 200 & {\em 10839} & 13000 $\pm$ 100 & {\em 13428} & 12400 $\pm$ 100 & {\em 12853} & 12300 $\pm$ 100 & {\em 12794} & 12600 $\pm$ 200 & {\em 12246} \\
T$_r$([O{\sc iii}]) ($K$)  & -- & {\em 11000} & -- & {\em 13600} & -- & {\em 13100} & -- & {\em 12900} & -- & {\em 12400} \\
$t^2$ & -- & {\em 0.004} & -- & {\em 0.001} & 0.001$^{+0.037}_{-0.001}$ & {\em 0.002 }& -- & {\em 0.003} & -- & {\em 0.002} \\
12+log(O$^{2+}$/H$^+$) & 7.84 $\pm$ 0.02 & {\em 7.93} & 7.96 $\pm$ 0.02 & {\em 7.93}  & 7.98 $\pm$ 0.02 & {\em 7.92 } & 7.87 $\pm$ 0.02 & {\em 7.78}  & 7.92 $\pm$ 0.02 & {\em 7.99}\\
12+log(Ne$^{2+}$/H$^+$) & 7.17 $\pm$ 0.04 & {\em 7.11} & 7.24 $\pm$ 0.03 & {\em 7.14}  & 7.33 $\pm$ 0.03 & {\em 7.16}   & 7.22 $\pm$ 0.04 & {\em 7.09 }  & 7.35 $\pm$ 0.04 & {\em 7.31} \\
12+log(Fe$^{2+}$/H$^+$) & 5.52 $\pm$ 0.09 & {\em 5.40}  & -- & 6.15 & {\em 5.48 } $\pm$ 0.05 & {\em 5.28 }  & 5.49 $\pm$ 0.08 & {\em 5.41 }  & 5.44 $\pm$ 0.08 & {\em 5.37} \\
\\
T([N{\sc ii}]) ($K$)  &  -- & {\em 11376 } & -- &{\em  12503}  & 14200 $\pm$ 800 & {\em 12134 } & -- & {\em 11995 } & 14000 $\pm$ 900 & {\em 11967}  \\
12+log(N$^+$/H$^+$) & 6.47 $\pm$ 0.07 &{\em  6.36}  & 5.67 $\pm$ 0.09 &{\em  5.63 } & 5.94 $\pm$ 0.06 &{\em  6.00}  & 6.15 $\pm$ 0.06 & {\em 6.23 } & 6.30 $\pm$ 0.07 & {\em 6.43} \\
\\
T([S{\sc ii}]) ($K$)  &  8500 $\pm$ 500 & {\em 11143}  & 12100 $\pm$ 1200 & {\em 12303}  & 10400 $\pm$ 700 &{\em  11858 } & 8800 $\pm$ 500 &{\em  11668}  & 8200 $\pm$ 600 & {\em 11695 } \\
12+log(S$^+$/H$^+$) & 6.19 $\pm$ 0.08 &{\em  5.88 } & 5.30 $\pm$ 0.10 &{\em  5.26 } & 5.69 $\pm$ 0.08 &{\em  5.49 } & 6.07 $\pm$ 0.08 &{\em  5.70}  & 5.95 $\pm$ 0.10 & {\em 5.58} \\
\\
T([S{\sc iii}]) ($K$) &  9700 $\pm$ 400 & {\em 11015 } & 12900 $\pm$ 700 & {\em 13366 } & 12600 $\pm$ 400 & {\em 12706}  & 14500 $\pm$ 800 & {\em 12589}  & 11300 $ \pm$ 500 &{\em  12218}  \\
12+log(S$^{2+}$/H$^+$) & 6.47 $\pm$ 0.06 &{\em  6.41 } & 6.13 $\pm$ 0.05 & {\em 6.14 }  & 6.14 $\pm$ 0.04 & {\em 6.15}  & 6.00 $\pm$ 0.06 & {\em 6.19}   & 6.31 $\pm$ 0.06 & {\em 6.25} \\
12+log(Ar$^{2+}$/H$^+$) & 5.95 $\pm$ 0.06 &{\em  5.78 } & 5.59 $\pm$ 0.06 &{\em  5.53}  & 5.66 $\pm$ 0.04 &{\em  5.61 }  & 5.49 $\pm$ 0.06 & {\em 5.51 }  & 5.78 $\pm$ 0.06 & {\em 5.72} \\
\\
log(He$^+$/H$^+$)  & -1.07 $\pm$ 0.03 & {\em -1.07}  & -1.08 $\pm$ 0.02 & {\em -1.08}    & -1.01 $\pm$ 0.03 & {\em -1.08}  & -1.08 $\pm$ 0.06 & {\em -1.08}  & -1.05 $\pm$ 0.05   &  {\em -0.99 }  \\
\hline
\end{tabular}
\end{center}
\end{minipage}
\end{table*}

\section{Discussion}

\subsection{Properties of the ionizing population}


In Table \ref{mod_res} we list the main observed properties of the young stellar population responsible for the 
ionization of the surrounding gas and their comparison with the same values as predicted by the corresponding 
tailor-made models. Column 2 lists
the age of the burst of continuous star formation adopted for each model in Myr. Column 3 gives 
the metallicity ($Z$) as derived from the relative oxygen
abundance measured in the gas-phase and column 4 the adopted $Z$ of the stellar ionizing cluster, 
which is in each case the  value closest to the oxygen abundance. 
Although the hypothesis of equal metallicity for the gas and the ionizing stars is not
well justified and the available metallicities in the stellar libraries are still poor to some extent, 
the differences in metallicity among models are not crucial to improve their accuracy. 
Column 5 in the table lists the number of ionizing photons, Q$_{obs}$(H), as derived from 
the observed luminosity of H$\alpha$ using the following expression (see, for example,
Osterbrock, 1989):

\begin{equation}
Q_{obs}(H) = 7.35 \cdot 10^{11} \cdot L(H\alpha)
\end{equation}

\noindent
with L(H$\alpha$) in erg $\cdot$ s$^{ -1}$. 

This Q$_{abs}$(H) should be compared to the the number of photons produced by the model cluster 
once divided by the dust absorption factor, $f_d$, predicted by the models, which is listed in column 6. 
The total number of ionizing photons produced by the ionizing cluster is then:

\begin{equation}
Q(H) = f_d \cdot Q_{obs}(H)
\end{equation}

\noindent where both Q(H) are expressed in photons $\cdot$ s$^{-1}$. Therefore, in order to be compared with the observations
we list in column 7 the ratio  $Q(H)/f_d$ predicted by the models.  Finally, columns 7 and 8 give the equivalent width of H$\beta$, EW(H$\beta$), measured on the
spectra once the underlying stellar populations have been removed. This is to be compared with the corresponding EW(H$\beta$) value given by the models,
which already takes into account the dust absorption factor and the nebular continuum emission. 
In the lower panel of the table we show in columns 2 and 3 the logarithm of
the ionization parameter, $\log U$, calculated using the following expression derived by  D\'\i az et al. (1991) from the emission lines
of [S{\sc ii}] and [S{\sc iii}]:

\begin{equation}
\log U = -1.68 \cdot \log \frac{I([SII] 6717,6731 \AA)}{I([SIII] 9069,9532 \AA)} - 2.99
\end{equation}

\noindent which offers a reliable estimation of the ionization parameter in nebulae with a
non-plane parallel geometry. This value is compared with the corresponding $\log U$ obtained from our models.
The logarithm of the filling factor ($\log\epsilon$) and the adopted inner
radius (R$_0$) of the ionized region are listed in columns 4 and 5 while columns 6 and 7 give the external radius (R$_S$) 
from the models as compared with the radius derived from the spatial profile of H$\alpha$
in the long slit. Finally, the last three columns give the dust-to-gas ratio in each model and the visual
extinction, A(V), as derived from the observed Balmer decrements and compared with the values obtained from the models.

As in the case of the emission lines, the agreement between the derived properties and those obtained from the models
is good with the exception of the outer radii which are much
larger in the observations by factors ranging from 2.2 (J1616) to 15.5 (J1540). Possible explanations for these discrepancies could be
either a very low density layer of diffuse gas in the outer regions of the ionized gas or lower filling factors as compared
with the values found in our models. We find as well some discrepancies between the extinction values found in
our models and those derived using Balmer decrements. This could be symptomatic of very complex structures
of dust in the ionized gas region, not easy to reproduce in the models.


\begin{table*}
\begin{minipage}{180mm}
\begin{center}
\scriptsize
\caption{Comparison between some properties obtained from the best tailor-made model
and derived from the observations for each object, including the age of the ionizing cluster, metallicity 
of the gas and the adopted metallicity of the cluster, dust absorption factor,
number of ionizing photons, equivalent width of H$\beta$, ionization parameter, filling factor, inner and outer
radii, dust-to-gas ratio and visual extinction.}
\label{mod_res}
\begin{tabular}{lcccccccccc}
\hline

ID Object & Age (Myr) & \multicolumn{2}{c}{Z}  &  log Q$_{obs}$(H) (s$^{-1}$) & Abs. factor $f_d$ & log Q(H)/$f_d$ &
\multicolumn{2}{c}{EW(H$\beta$) (\AA)} \\
&{\em Model} & Observed  &  {\em Model} &  Observed  &  {\em Model} & {\em Model} & Observed  &  {\em Model} \\ 
\hline
J0021  &   {\em   8.9 }   &    0.0055 $\pm$ 0.006 &  {\em   0.004 }   &   54.00 $\pm$ 0.01  & {\em 1.763 }  & {\em 53.98}  &        110 $\pm$ 2  &   {\em 112}     \\
J0032  &   {\em  8.9 }    &    0.0037 $\pm$ 0.003 &  {\em   0.004 }   &   52.17 $\pm$ 0.01 & {\em 1.508}   & {\em 52.22}  &        119 $\pm$ 2  &     {\em 121 }    \\
J1455  &   {\em   4.5 }   &    0.0038 $\pm$ 0.004 &  {\em   0.004 }   &   52.70 $\pm$ 0.01 & {\em 2.609 }   & {\em 52.58 } &        146 $\pm$ 2  &    {\em  149  }   \\
J1509  &   {\em  7.0  }   &    0.0068 $\pm$ 0.007 &  {\em   0.008}    &  53.13 $\pm$ 0.01 &  {\em 1.500  }  & {\em 53.13 } &        135 $\pm$ 2  &     {\em 132 }    \\
J1528  &   {\em  5.5 }    &    0.0065 $\pm$ 0.007 &  {\em   0.008 }   &   53.50 $\pm$ 0.01 &  {\em 1.581 }  &  {\em 53.48 } &        191 $\pm$ 2  &     {\em 182}     \\
J1540  &   {\em  15.8 }   &    0.0045 $\pm$ 0.005 &  {\em   0.004 }   &  51.51 $\pm$ 0.01 & {\em 1.027 }  &  {\em 51.69 } &        124 $\pm$ 2  &     {\em 125}     \\
J1616  &   {\em  10.0}    &    0.0051 $\pm$ 0.004 &  {\em   0.004}    &  50.80 $\pm$ 0.01 & {\em 2.107}  &  {\em 50.78 } &        83 $\pm$ 2   &     {\em 97 }     \\
J1624  &   {\em  7.9 }    &    0.0049 $\pm$ 0.003 &  {\em   0.004 }   &  53.34 $\pm$ 0.01 &  {\em 1.969}  & {\em 53.41 } &        107 $\pm$ 2  &     {\em 111}     \\
J1657  &   {\em  7.9}     &    0.0043 $\pm$ 0.004 &   {\em  0.004}    &  52.60 $\pm$ 0.01 &   {\em 1.621}  &  {\em 52.59}  &        132 $\pm$ 2  &     {\em 127 }    \\
J1729  &   {\em  7.9 }    &    0.0053 $\pm$ 0.005 &   {\em  0.004 }   &   52.43 $\pm$ 0.01  & {\em 1.928}  &  {\em 52.30 } &        126 $\pm$ 2  &    {\em  119}     \\
\hline
\hline
ID Object & \multicolumn{2}{c}{log U} & Fil. fac. $\log\epsilon$ & R$_0$ (pc) & \multicolumn{2}{c}{R$_{S}$  (pc)} & Dust-to-gas & \multicolumn{2}{c}{A(V)} \\
& Observed & {\em Model} & {\em Model} & {\em Model} & Observed  &  {\em Model} &  & Observed  &  {\em Model} \\
\hline
J0021   & -2.71 $\pm$ 0.04 & {\em -2.66}  & {\em -2.99}  &  {\em 22.7}   &    5100 $\pm$ 200  &   {\em 1541}  &     {\em 8.76 $\cdot$ 10$^{-3}$} &   1.09 $\pm$ 0.02  &    {\em 0.37}   \\
J0032   & -2.41 $\pm$ 0.05 & {\em -2.61}  & {\em -1.94}  &  {\em 28.5}   &    1400 $\pm$ 200  &    {\em 177}    &    {\em 5.24 $\cdot$ 10$^{-3}$}  &     0.24 $\pm$ 0.02  &    {\em 0.27 }  \\
J1455   & -2.33 $\pm$ 0.03 & {\em -2.29}  & {\em -1.95}  &  {\em 2.9}   &    2900 $\pm$ 600  &    {\em 240}  &      {\em 9.24 $\cdot$ 10$^{-3}$}  & 0.28 $\pm$ 0.02  &   {\em  0.63}   \\
J1509   & -2.40 $\pm$ 0.06 & {\em -2.52}  & {\em -2.30}  &  {\em 2.9}   &    3900 $\pm$ 100   &    {\em 449}  &      {\em 4.19 $\cdot$ 10$^{-3}$}  &0.15 $\pm$ 0.02   &    {\em 0.25 }  \\
J1528   & -2.60 $\pm$ 0.06 & {\em -2.77}  & {\em -2.84}  &  {\em 181.5}  &    3900 $\pm$ 200  &   {\em  938 }   &     {\em 10.58 $\cdot$ 10$^{-3}$}  & 0.09 $\pm$ 0.02   &   {\em  0.30 }  \\
J1540   & -2.63 $\pm$ 0.04 & {\em -2.84}  & {\em -1.62}  &  {\em 67.7}   &    1200 $\pm$ 100  &    {\em 101}   &      {\em 1.83 $\cdot$ 10$^{-5}$}  & 0.47 $\pm$ 0.02  &    {\em 0.00}   \\
J1616   & -1.95 $\pm$ 0.04 & {\em -2.13}  & {\em -0.67}  &  {\em 5.6}   &     340 $\pm$ 70  &    {\em 22}   &      {\em 5.46 $\cdot$ 10$^{-3}$}  & 0.04 $\pm$ 0.02 &    {\em 0.50 }  \\
J1624   & -2.48 $\pm$ 0.04 & {\em -2.49}  & {\em -2.52}  &  {\em 4.8}   &    2900 $\pm$ 200  &    {\em 684}  &    {\em 7.44 $\cdot$ 10$^{-3}$}  &  1.12 $\pm$ 0.02   &    {\em 0.42 }  \\
J1657   & -2.78 $\pm$ 0.05 & {\em -2.84}  & {\em -2.50}  &  {\em 15.9}   &    4000 $\pm$ 1000   &    {\em 374}  &    {\em 10.90 $\cdot$ 10$^{-3}$}  &   0.11 $\pm$ 0.02  &    {\em 0.32 }  \\
J1729   & -2.20 $\pm$ 0.04 & {\em -2.49}  & {\em -1.92}  &  {\em 4.4 }  &    1500 $\pm$ 100   &    {\em 182}  &     {\em 6.57 $\cdot$ 10$^{-3}$}  & 0.06 $\pm$ 0.02  &    {\em 0.37 }  \\
\hline
\hline
\end{tabular}
\end{center}
\end{minipage}
\end{table*}


\begin{figure*}
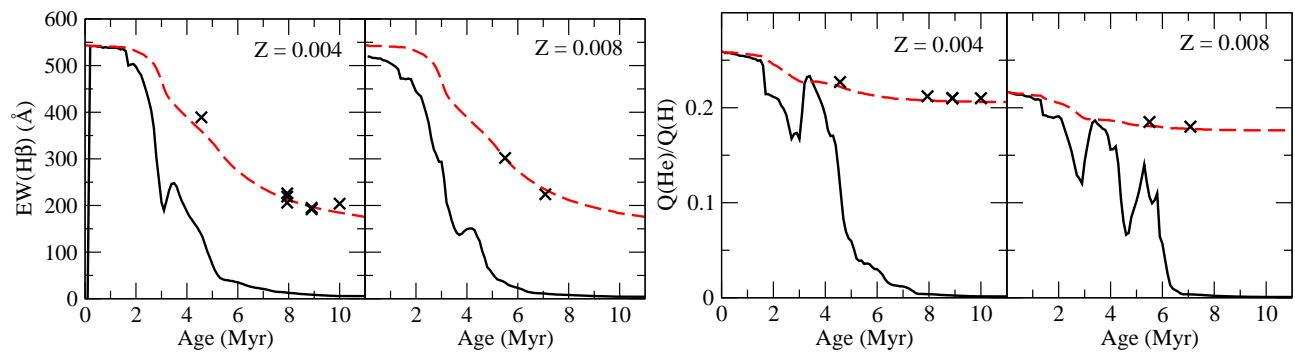


\begin{minipage}{170mm}
\centerline{
\psfig{figure=mn-ghiimod.fg05l.eps,width=8.5cm,clip=}
\psfig{figure=mn-ghiimod.fg05r.eps,width=8.5cm,clip=}}
\caption{Left panels: relation between the equivalent width of H$\beta$ and the age of the corresponding 
synthesis 
population model for metallicities of 0.004 and 0.008 as predicted by Starburst99. 
Crosses represent the observed values measured
on the young stellar continua and 
multiplied by the corresponding absorption factor obtained in each tailor-made model.
The assigned age to each object is that predicted by the same models. Therefore, the
J1540 point does not appear as the fitted age is older than 10 Myr.
Right panels: relation between
the ratio of helium to hydrogen ionizing photons and the age of the ionizing cluster. In all panels, the black
solid line represents an instantaneous star formation and the red dashed line, a continuous star formation
as predicted by Starburst99. 
As in the previous case, crosses represent the values obtained from the models for each object, with the
exception of J1540.}
\label{ewhb}
\end{minipage}
\end{figure*} 

Regarding the age of the burst, the equivalent width of an intense Balmer emission line, like H$\beta$,
is generally associated to the age and properties of the ionizing population. It has long been accepted that
the low values of EW(H$\beta$) found in starburst galaxies as compared with the predictions
from Population Synthesis Models are due to the presence of an
underlying stellar population that enhances the continuum light without significantly contributing to the ionization of the gas.
Nevertheless, as can be seen in Table \ref{fitting} the subtraction of this population from the integrated spectra  
does not imply significant corrections to EW(H$\beta$) (between 1.5 \% in the case of J1540 and J1729
and 24 \% in the case of J0032). 

However, the excitation conditions found in these objects,
are only reachable with the energy
supplied by very young and massive O stars, whose
associated EW(H$\beta$) is four or five times larger than the observed values.
This difference can be partially explained by a certain 
dust absorption of UV photons.
The dust absorption factors obtained in our
models range between 1.5 and 2, as we can see in Table \ref{mod_res}.
These help to put the modeled EW(H$\beta$) values in agreement with the observed ones.

In the left panels of Figure \ref{ewhb} we show the evolution of EW(H$\beta$) with time for the two metallicities closest 
to those in our sample for both instantaneous (black solid line) and continuous (dashed red line)
star formation laws as obtained from Starburst99 models.
In the right panels of the same figure we show the evolution of the ratio of helium to hydrogen ionizing photons, Q(He)/Q(H)
according to the same Starburst99 models.
This ratio, that can be considered a good estimate of the ionizing power of the cluster, falls dramatically for an
instantaneous star formation law once the WR phase has finished ($\sim$ 5 Myr), while it maintains a
constant high value in the case of a continuous star formation. 
We can compare these diagrams with 
the observed EW(H$\beta$), once the underlying stellar population has been removed and the dust absorption factor has been taken into account. 
They appear as crosses in Figure \ref{ewhb} and lie in a range which
corresponds to the asymptotic value of the continuous star formation line.
In the case of the evolution of Q(He)/Q(H), crosses represent the values derived from the tailor-made models.

Although it is true that both EW(H$\beta$) and Q(He)/Q(H) could be consistent with an instantaneous
burst at present in the WR phase, statistically it is not probable that all the objects of the sample present such
similar properties due to their being in the same evolutionary stage. At the same time, all the calculated photoionization
models in our work require a great amount of O and B stars younger than the WR phase in order
to reach the ionization conditions responsible for the observed emission lines.
We have therefore assumed for all the studied objects a continuous star formation history
for the ionizing source, with a duration longer in all cases than 5 Myr.


\begin{figure*}
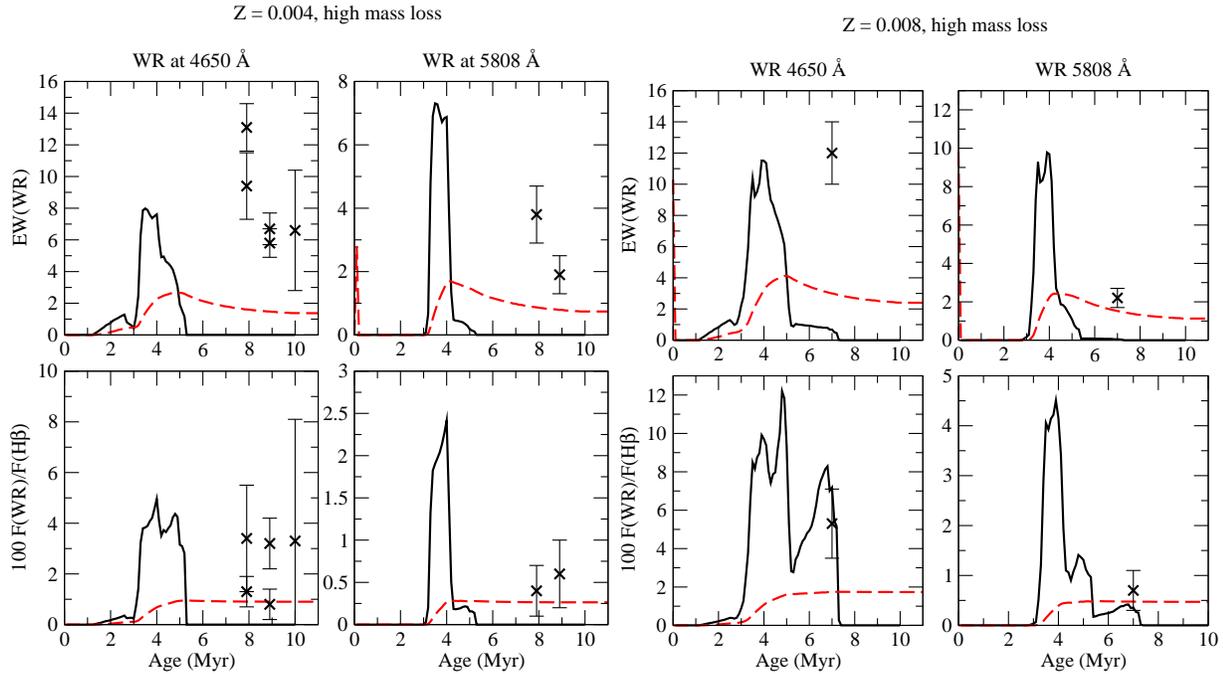

\begin{minipage}{170mm}
\centerline{
\psfig{figure=mn-ghiimod.fg06l.eps,width=8cm,clip=}
\psfig{figure=mn-ghiimod.fg06r.eps,width=8cm,clip=}}
\caption{Relation between the WR equivalent widths and relative intensities of both the blue and the red bumps 
as a function of the age of the cluster, according to Starburst99 model predictions, 
at left for a metallicity of
Z = 0.004 and at right for Z = 0.008. In all panels, black solid lines represent the instantaneous star formation
model and red dashed lines the continuous star formation one. Crosses represent the corrected values
derived for each galaxy and listed in Table 3 }
\label{wrf}
\end{minipage}
\end{figure*} 


\begin{figure*}
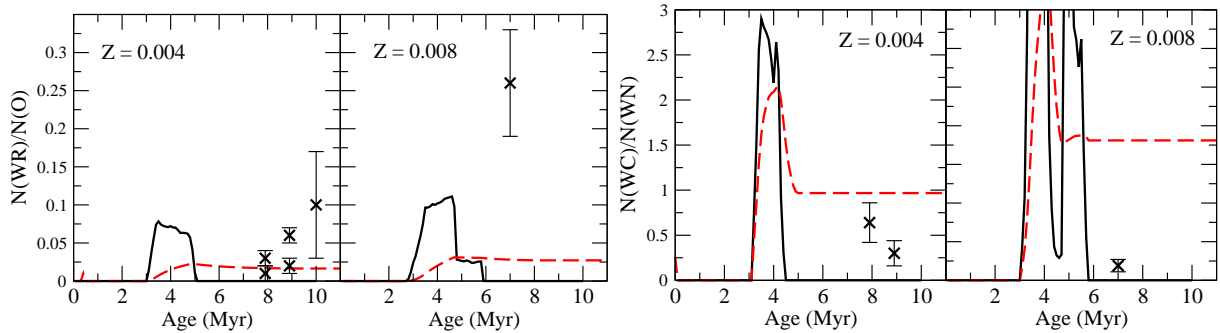

\begin{minipage}{170mm}
\centerline{
\psfig{figure=mn-ghiimod.fg07l.eps,width=8cm,clip=}
\psfig{figure=mn-ghiimod.fg07r.eps,width=8cm,clip=}}
\caption{Left panels: time evolution of the ratio of WR to O stars according to Starburst99 predictions. Right panel: same but for the ratio of WC to WN stars. In both cases, results for two metallicities, Z= 0.004 and Z=0.008, are shown. Solid lines represent instantaneous star
formation models and dashed line continuous star formation ones. Crosses represent the derived numbers for each galaxy
taking into account the measured luminosities of the bumps as compared with those predicted by the Starburst99 models at the
predicted age. The number of O stars has been derived using the same procedure for L(H$\alpha$).}
\label{wrstars}
\end{minipage}
\end{figure*} 

\subsection{Derived properties of the WR population}

Seven out of the ten observed galaxies have detectable WR features in their spectra. 
Five of them were already included in the catalog by Kniazev et al. (2004; SHOC) with a flag 
indicating the detection of both the blue and red bumps from WR stars.  In our spectra, the red bump is observed in only three galaxies.  The SHOC catalog, however, does not provide actual measurements for the features. In a more recent paper, 
Brinchmann et al. (2008) list only the "blue bump" luminosities and equivalent widths 
of three galaxies from the present work: J002101 (SHOC 11), J003218 (SHOC 22) 
and J162410 (SHOC 536). The direct comparison between the equivalent widths of 
the blue bump measured by these authors inside the 3-arcsec fiber spectra of the SDSS and our
long-slit observations yields discrepancy factors for J0021, J0032 and J1624 of
1.7, 2.0 and 1.1, respectively, larger in our WHT observations. This could be a 
consequence of the smaller aperture of
our long-slit observations which includes a lower contribution to the continua from extended 
ionized gas and hence produce larger equivalent widths.


Table \ref{wr_numbers} shows the extinction corrected luminosities of the measured WR bumps. 
We show in the same table the number of O and WR stars as derived from the extinction corrected 
H$\alpha$ and WR blue bump luminosities, respectively.  In the case of O stars, we have taken into
account as well the dust absorption factor, $f_d$.
The total number of stars have been estimated by comparison with the properties of the Starburst99 models at the
age derived from the photoionization models described above. 
For those galaxies with a red bump 
detection  we have used the same libraries to derive the ratio between WC and WN
stars, which is also listed in Table \ref{wr_numbers}.


\begin{table*}
\begin{minipage}{180mm}
\centering
\caption[]{Luminosities of the Wolf-Rayet features measured for the
sample of HII galaxies at 4650 {\AA} (blue bump, bb) and at 5808 {\AA} (red bump, rb), derived number of
WR stars, ratios of WR to O stars and, for those galaxies with a detection of the red bump, of
WC to WN stars.}
\label{wr_numbers}
\small
\begin{tabular} {c c c c c c}

\hline
\hline
 \multicolumn{1}{c}{Object  ID}  &       log L(bb)         &      log L(rb)         &    N(WR) &    WR/O &    WC/WN \\
                                 & (erg $\cdot$ s$^{-1}$)  & (erg $\cdot$ s$^{-1}$) &          &         &          \\
\hline

J0021   & 39.84$\pm$0.09 & --             & 2500$\pm$600 & 0.02$\pm$0.01 & --            \\
J0032   & 38.53$\pm$0.08 & 37.79$\pm$0.16 &  150$\pm$30  & 0.06$\pm$0.01 & 0.30$\pm$0.14 \\
J1509   & 39.73$\pm$0.09 & 38.84$\pm$0.11 & 1750$\pm$400 & 0.11$\pm$0.02 & 0.19$\pm$0.08 \\
J1540   & 38.18$\pm$0.06 & --             &   60$\pm$15  & 0.26$\pm$0.07 & --            \\
J1616   & 37.32$\pm$0.23 & --             &    9$\pm$6   & 0.10$\pm$0.07 & --            \\
J1624   & 39.46$\pm$0.08 & 38.92$\pm$0.12 & 1070$\pm$200 & 0.03$\pm$0.01 & 0.64$\pm$0.22 \\
J1729   & 38.92$\pm$0.12 & --             &  300$\pm$100 & 0.01$\pm$0.01 & --            \\
\hline

\end{tabular}
\end{minipage}
\end{table*}


We can compare the
relative intensities and equivalent widths of the WR bumps detected in our objects
with the predictions of Starburst99 populations synthesis models
as a function of the metallicity, age and star formation law of the cluster.
To make this comparison we have taken the corrected values listed in Table 3: in the case of EWs
from the effect of the underlying stellar populations and the contribution of the nebular continuum and
in the case of the relative intensities from the effect of the dust absorption factor to the
hydrogen recombination lines.
In Figure \ref{wrf} we show the predicted equivalent widths and intensities relative to H$\beta$ 
of both the blue and red bumps as a function of the age of the cluster for metallicities Z = 0.004 (0.2$\cdot$Z$_{\odot}$), left panels, and Z = 0.008 (0.4$\cdot$Z$_{\odot}$), right panels.
The black solid lines represent the evolution of
an instantaneous burst, while the red dashed lines do for a continuous star formation
history with a constant star formation rate. 
Crosses represent the corrected values as listed in Table 3 with their corresponding error bars.
As we can see, the models predict the brightest and most prominent features
for the highest metallicity, agreeing with the stellar model atmospheres for WR stars (Crowther, 2007
and references therein).
At the same time, we can appreciate that in the instantaneous star formation case the WR features appear
during a time interval between 2 and 5 Myr
reaching higher intensities than in the case of continuous star formation. In this latter scenario, the WR features
appear at the same age and, despite reaching lower intensities, they converge to a non-zero value 
at older ages. For an average metallicity between the two cases of study these values are
EW $\sim$ 2 {\AA} and 0.013$\cdot$I(H$\beta$) for the blue bump and EW $\sim$ 1 {\AA} 
and 0.004$\cdot$I(H$\beta$) for the red one.
By comparing with the measured relative intensities and equivalent widths in our
sample objects we can see that all of them are larger than the Starburst99 model predictions
for a constant star formation law and,
in some of them, like J1509 and J1540, even larger than the values predicted for an
instantaneous burst.


\begin{figure*}
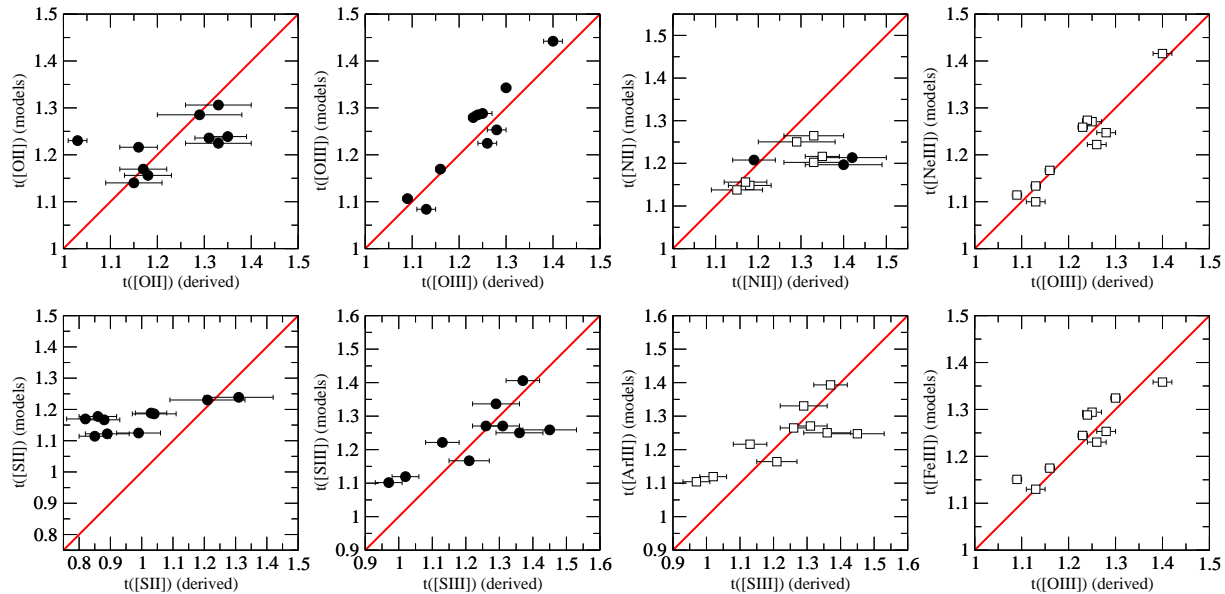

\begin{minipage}{170mm}
\centerline{
\psfig{figure=mn-ghiimod.fg08l.eps,width=8cm,clip=}
\psfig{figure=mn-ghiimod.fg08r.eps,width=8cm,clip=}}
\caption{Comparison between the electron temperatures (in units of 10$^4$ K) as derived from the observations
and from the tailor-made photoionization models. Black circles represent a comparison between a temperature
directly derived in the observations, while white squares represent a temperature of
another ion, assumed to be similar in the thermal structure.  }
\label{temps}
\end{minipage}
\end{figure*} 

A great deal of work has been done in order to reconcile this disagreement 
between the observed luminosities of
the WR bumps detected in the integrated spectra of galaxies and the 
model stellar atmospheres of WR stars and 
evolutionary synthesis models of clusters. Suggested hypotheses include
the existence of a binary channel (Eldridge et al., 2007) or the consideration of rotation in the stellar evolutionary 
tracks, which reduces the mass required  to form WR stars (Meynet \& Maeder, 2005).
Nevertheless, some other reasons can be explored 
such as geometrical effects.
If we consider the whole 
burst of star formation and we assume a continuous star formation law, we
must compare the luminosities of the WR bumps with the intensities of the
Balmer lines and the continuum coming from the whole burst. This approach was already
used by P\'erez-Montero \& D\'\i az (2007) to match the observed fluxes of the WR features
in Mrk 209. On the contrary, if only the
ionizing stellar population producing the WR stars is considered, 
only the burst region where the WR population concentrates must be analyzed.
This could explain as well the lack of WR emission in the other three galaxies with
an assumed continuous star formation history.
In the case of our sample, no aperture correction factors have been taken into account
either for the H$\alpha$ luminosities or the continuum contribution at the
WR bumps wavelength, due to the compact nature of the objects.
Indeed, the discrepancy factors between our H$\alpha$ luminosities and those 
measured in the SDSS catalogue using a 3 arcsec fiber is not larger than 1.5 in
any case and even smaller than 1 in two of them (J1528 and J1729).
Therefore, the aperture factor does not seem to be the cause of the abnormally high
values found for the EW and the relative intensities of the WR features.


The analysis of the relative number of WR to O stars and of the number of 
WC to WN stars leads to very similar conclusions. In all cases the derived
ratio of the number of WR to O stars is much higher than the asymptotic value reached 
with a continuous star formation, which is 0.022 for a metallicity in between the two
studied cases. For J1540, the obtained value, 0.44, is even
much higher than the value expected for an instantaneous burst. As in the case
of the luminosities of the WR features, we can reconcile the derived values enhancing
the number of O stars on the basis that a non-negligible flux of the radiation
from these stars is obscured by internal extinction. As in the previous case,
probably geometrical effects can be determinant. Regarding the ratio between
WC and WN stars, the derived values are lower in all cases than expected.
This is due to the fact that the luminosities of the red bump predicted by the models are in much better
agreement with the observed values than those of the blue bump and, hence, the relative number
of WC stars is lower.

\subsection{Physical conditions and ionic abundances of the gas through tailor-made models}

The correct determination of ionic abundances in ionized gaseous nebulae using
collisional emission lines versus recombination lines has become a matter of debate in the last years 
due to the disagreements
predicted by photoionization models in the high metallicity regime ({\em e.g.} Charlot \& Longhetti, 2001) 
or the discrepancies found in local nebulae  ({\em e.g.} Peimbert et al., 2007 and references therein).

Many times, these discrepancies are due to the simple assumptions made about the
ionization structure of the nebulae. Indeed, the only electron temperature measured in many cases
is t([O{\sc iii}]) because the emission line of [O{\sc iii}] at 4363 {\AA} is the brightest
auroral line. This temperature is assumed to be representative of the high excitation zone and the 
low excitation zone temperature is obtained under very simple assumptions. This can produce large deviations in 
high metallicity nebulae, where the low excitation ions have a high relative
weight in the total abundance of a specific species (P\'erez-Montero \& D\'\i az, 2005). 
One way to overcome this issue is to try to trace as precisely as possible the inner thermal
structure of the gas. 


\begin{figure*}
\begin{minipage}{170mm}
\centerline{
\psfig{figure=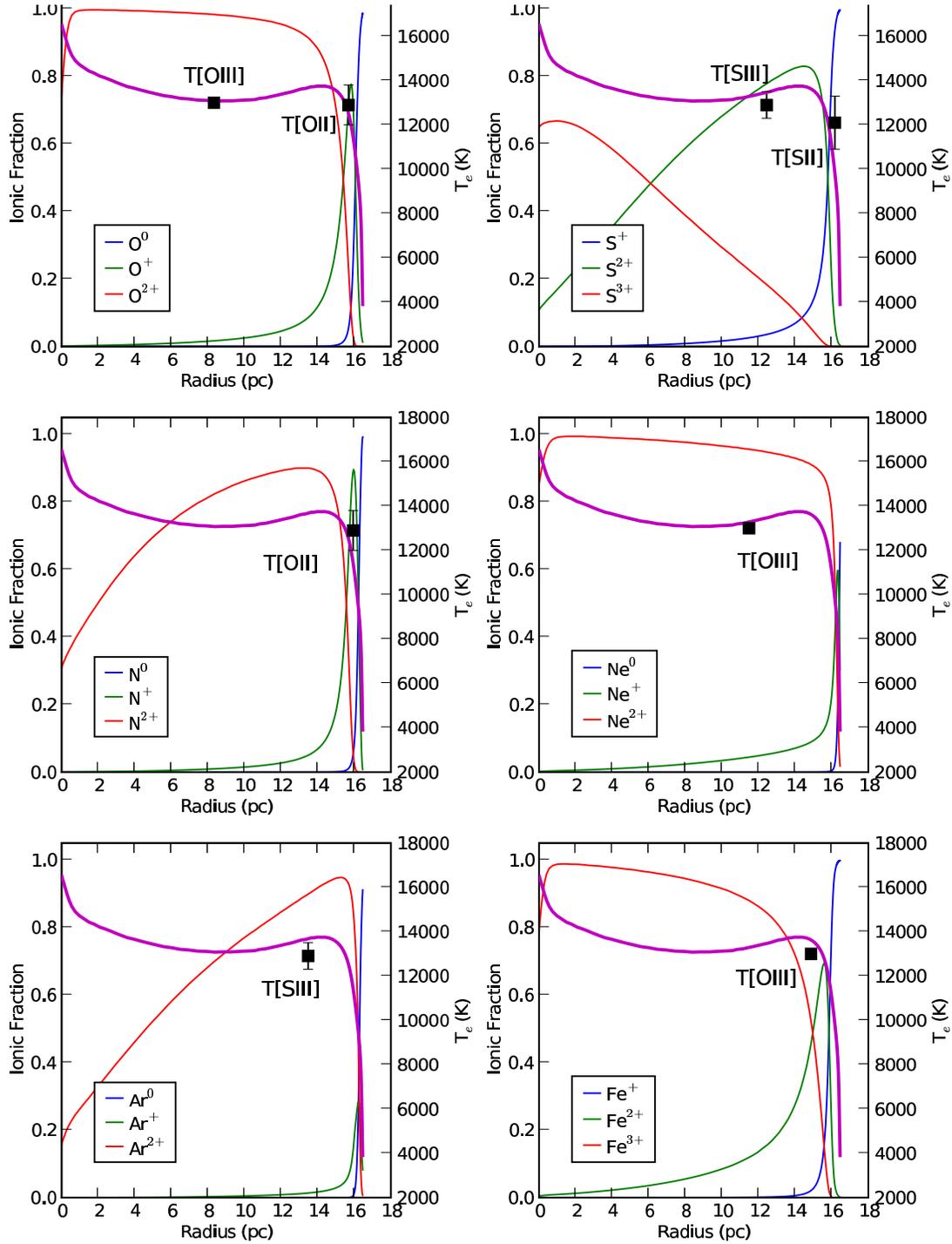,width=15cm,clip=}}
\caption{Radial profile of the electron temperature, in violet solid line, and the relative abundance of
different ionic species (from left to right and up to down, O, S, N, Ne, Ar and Fe)  as calculated in 
the best photoionization tailor-made model of the galaxy J1616. We show as well the derived 
electron temperatures of oxygen and sulphur as black squares in the position of the 
average abundance of the corresponding ions.} 
\label{J1616_temps}
\end{minipage}
\end{figure*} 


\begin{figure*}
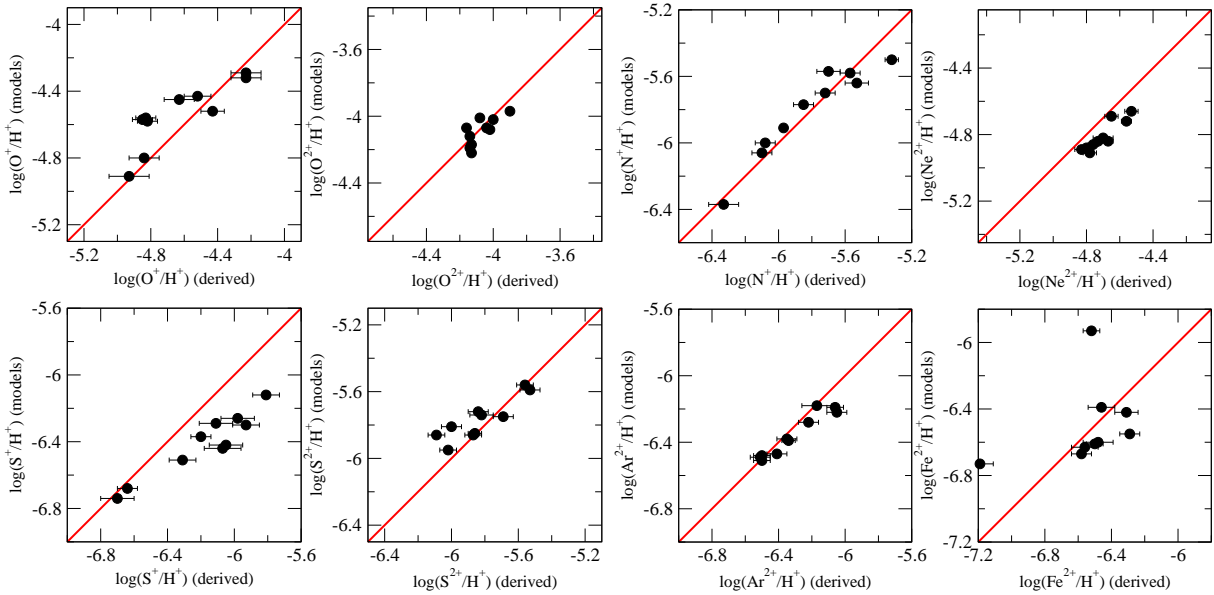

\begin{minipage}{170mm}
\centerline{
\psfig{figure=mn-ghiimod.fg10l.eps,width=8cm,clip=}
\psfig{figure=mn-ghiimod.fg10r.eps,width=8cm,clip=}}
\caption{Comparison between ionic abundances as derived from the observations using the direct
method and the predictions of tailor-made photoionization models.}
\label{ions}
\end{minipage}
\end{figure*} 

In Figure \ref{temps} we show a comparison between derived and modeled temperatures
for the objects of our sample, for which at least four electron temperatures have been measured.
In the case of ions whose temperatures can not be derived due to the lack of observable lines in the optical,  
another line temperature, assumed to be representative of the zone where these iones are formed, has been used: 
t([S{\sc iii}]) in the case of [Ar{\sc iii}],  t([O{\sc iii}]) in the case of [Ne{\sc iii}] and [Fe{\sc iii}] and 
t([O{\sc ii}]) in the case of [N{\sc ii}] in the objects where this was not available.
As we can see, the agreement for t([O{\sc iii}]) is excellent, mostly due to the input 
conditions in the models to fit the auroral line of [O{\sc iii}]. Regarding t([O{\sc ii}]) and
t([S{\sc iii}]), the agreement is still good, although slightly worse that in the previous case.
Finally, the largest deviations are found for the electron temperature of [S{\sc ii}].
Except in the cases of J1455 and J1616, models predict much higher
temperatures than observed. This is not surprising if we take into account that the
emission of [S{\sc ii}] is often found as well in the diffuse medium ({\em e.g.} Hunter \& Gallagher, 1990) and therefore we
are probably tracing in our observations a non-negligible portion of colder and less dense
gas, which has not been considered in our models. 
This is compatible with the significatively lower radii found in our models in comparison with
the values obtained from the spatial profile of H$\alpha$. Perhaps, this could be overcome if
we considered in
our models a low-density region in the outer parts of the nebula where lines from
low excitation ions like [S{\sc ii}] were produced.  
In the case of the three objects with a measurement of
the electron temperature of [N{\sc ii}], the agreement is rather good in the case of J0021, while it
is sensibly underestimated by the models in the case of J1624 and J1729. 
This temperature is assumed to be equal to t([O{\sc ii}]) in objects without a 
measurement of the auroral line of [N{\sc ii}] at 5755 {\AA}. As we can see in the corresponding
panel, this assumption seems to work for most of the objects, except in the cases of
J0032 and J1455. For these objects, we can not rule out some geometrical effect not considered
in our photoionization models.

For the rest of assumptions, as
t([Ar{\sc iii}]) $\sim$ t([S{\sc iii}]) or t([Ne{\sc iii}]) $\sim$ t([Fe{\sc iii}]) $\sim$ t([O{\sc iii}]),
we can see that all of them work fairly well.

In Figure \ref{J1616_temps} we illustrate the good agreement found between models and
observations for the galaxy J1616. In each panel, in red, green and blue solid lines, 
the radial distribution of the ionic fractions of the different elements 
as computed by the model (from left to right and from up to down: O, S, N, Ne, Ar and Fe) are plotted. 
The magenta thick solid lines show the radial distribution of the electron temperature predicted
by the same model. Finally, black squares show the electron temperatures obtained from the
observations. We have associated these points to the average zone where the corresponding ion is formed.  
In the cases of Ne$^{2+}$ and Fe$^{2+}$, t([O{\sc iii}]) is shown; for Ar$^{2+}$, t([S{\sc iii}]) is plotted 
and in the cases in which no measurements of t([N{\sc ii}]) are available, this is represented by t([O{\sc ii}]).

In Figure \ref{ions}, we show the comparison between the relative ionic abundances, derived
using the corresponding measured electron temperatures and the observed emission-line intensities,
and the same abundances as predicted by our photoionization models. 
We can see that the agreement is somewhat better for high excitation ions, like O$^{2+}$, than for low excitation ones, like O$^+$ and S$^+$. At the
same time these latter ions present abundances spanning a wider range than the former. 
This might indicate that geometrical effects ({\em e.g.} density variations,
matter bounded nebulae, etc...) affect the diagnostics of the gas in the low excitation
region, hence a simple scheme can not be adopted for its analysis.
In the case of our sample of HII galaxies, these differences are not appreciable due to their
low metallicities and high excitation conditions, so the inner parts have a large weight
on their average conditions. On the contrary, the analysis of disk HII regions and high metallicity objects
are tied to a much larger uncertainty if these effects are not taken into account using
as many indicators of the physical conditions of the gas as possible.


\begin{figure*}
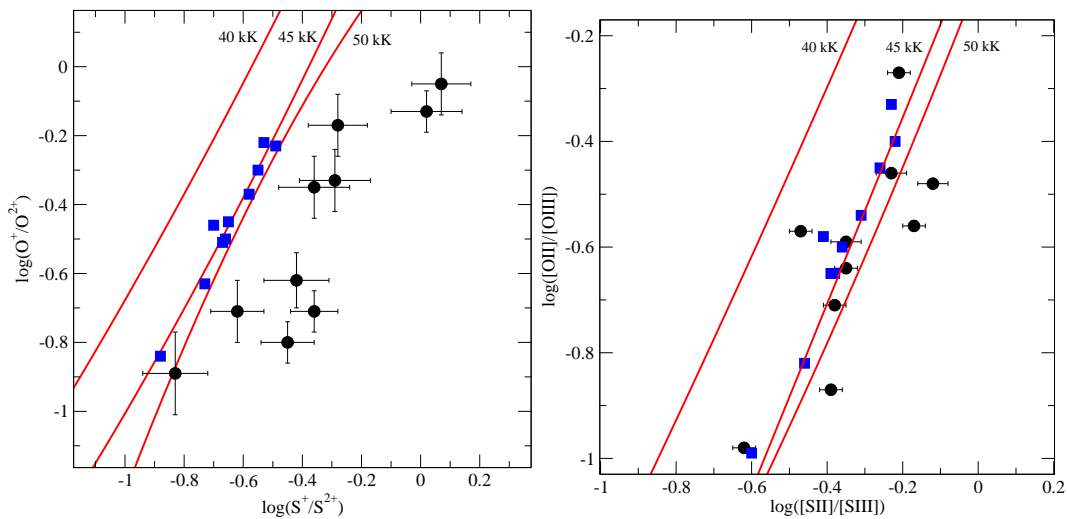

\begin{minipage}{170mm}
\centerline{
\psfig{figure=mn-ghiimod.fg11l.eps,width=7cm,clip=}
\psfig{figure=mn-ghiimod.fg11r.eps,width=7cm,clip=}}
\caption{At left, relation between the ratio of ionic abundances S$^+$/S$^{2+}$ and O$^+$/O$^{2+}$, whose
quotient is the $\eta$ parameter. Black dots with error bars represent the modeled objects with the corresponding
derived ratios, while blue squares are the values derived from the models. Red solid lines represent
the fits to a sequence of photoionization models with the same effective temperature. The right panel
represents the relation between the ratios of [SII]/[SIII] and [OII]/[OIII]. Symbols are the same as
in left panel.}
\label{eta}
\end{minipage}
\end{figure*} 

\subsection{The hardening of the radiation}

The stellar effective temperature has been traditionally the functional parameter most
difficult to obtain. V\'\i lchez \& Pagel (1988) defined the so-called softness
parameter, $\eta$, which allows to estimate the hardening of the radiation field by means
of the ratio of the ionic abundances of oxygen and sulphur, derived from the optical spectrum.

\begin{equation}
\eta = \frac{O^+/O^{2+}}{S^+/S^{2+}}
\end{equation}

The same ratio, defined using the intensities of the corresponding emission lines, can also be
used to figure out the scale of equivalent effective temperature:

\begin{equation}
\eta' = \frac{[OII] 3727/ [OIII] 4959,5007}{[SII] 6717,6731/ [SIII] 9069,9532}
\end{equation}

Both expressions agree in finding that HII galaxies
present very high equivalent effective temperatures as compared to other HII regions of higher
metallicity. 
Nevertheless, there is a disagreement
between the position of these objects in diagnostic diagrams and the scale of temperatures of
the stellar atmospheres of O stars. HII galaxies usually lie in a region of the diagrams which
corresponds to temperatures even higher than the maximum value of model atmospheres, 
which is 50 kK. The cause has
been sometimes attributed to some extra source of heating in HII galaxies  ({\em e.g}. Stasi\'{n}ska \&
Schaerer, 1999). 

The left panel of Figure \ref{eta} shows the logarithmic ionic fractions  O$^+$/O$^{2+}$ vs. S$^+$/S$^{2+}$. 
The different lines of constant slope 1 in this diagram correspond to different values of the $\eta$ parameter and hence of 
ionizing radiation temperature. The values derived from actual measurements of our observed objects are plotted as black solid circles while blue squares correspond to the values derived from the computed models. 
Red solid lines represent polynomial fits to different photoionization
models computed using as ionizing source WM-Basic stellar atmospheres (Pauldrach et al., 2001)
for different values of the effective temperature. These model sequences can help to
establish the scale of the equivalent effective temperature. We can see that the models are not able to reproduce the high values of log(S$^+$/S$^{2+}$), and in some cases log(O$^+$/O$^{2+}$) found in our objects that, in fact, would correspond to extremely high values of equivalent effective temperatures.

This disagreement disappears when the emission lines instead of the actual ionic ratios are considered. This is due to the fact that emission-line intensities are well fitted in all the tailor-made models while this is not the case for electron temperatures (see Figure \ref{temps}), specially for T([SII]) which is, in general overestimated by the models leading to an underestimate of the S$^+$/H$^+$ abundance ratio. Again, the disagreement could be caused by the geometry of the diffuse gas in these HII regions, which is not adequately modeled, thus avoiding the need for an extra source of heating to explain the apparently high stellar effective temperatures in these objects. At the same time, this probably means that the calibration of the effective temperature is not unique but in fact sensitive to the object class involved, through the different geometries of the ionized gas.

\section{Summary and conclusions}

We have studied the underlying and ionizing stellar populations
and have modeled the properties of the emitting gas in a sample of HII galaxies.
This sample was observed (H06 and H08) using double-arm
spectrographs, which allow the simultaneous analysis of the
entire spectral range from [OII] 3727 \AA\ to [SIII] 9532 \AA\ providing the derivation of the electronic temperatures of, at
least, four ions in all the cases: O$^+$, O$^{2+}$, S$^+$, and S$^{2+}$ and hence a precise determination of the ionization structure of the gas.

The fitting of the blue part of the continuum was made
using the program STARLIGHT and the SEDs computed from the Starburst99 stellar population synthesis models. This approach was used to study in a consistent
manner the differences between the stellar populations in our
sample of galaxies. The resulting distribution of ages and masses from
this fitting revealed the presence of various stellar populations.
An underlying stellar population of about several Gyr dominates the stellar
mass, while the blue to visual light is in all cases dominated by young stellar populations with ages younger than 10 Myr. 
The presence of the older population can affect the reddening determination but its contribution to the measured EW(H$\beta$) is only about 10\% for most of the sample objects, being around 25\% in the case of J0032 and almost zero for J1616 and J1729.
In fact, regarding the subtraction of Balmer absorption features from the underlying stellar populations,  we did not find that the spectral fitting technique constitute a substantial improvement over the simultaneous fitting of absorption and emission components as performed in previous works (H06 and H08).

We analysed the Wolf-Rayet clusters detected in seven of
the ten studied galaxies through the WR broad features seen in their spectra. 
The fitting of the stellar population and the modeling of the ionized gas allows to
correct them for the contribution of the underlying population and the nebular emission to the
continuum and for the dust absorption factor in the case of the emission of the Balmer lines.
The comparison between the corrected
fluxes and equivalent widths of the WR blue bump
with the predictions from Starburst99 models at the corresponding
metallicity leads to the estimate of the number
of WR stars which ranges from 9$\pm$6 to 2500$\pm$600 depending on the galaxy. In three of them, also the red bump
was measured allowing to estimate the WC to WN number ratio, Nevertheless,
the fluxes of both the blue and red bumps relative to H$\beta$ are larger than the values
predicted by the models in almost all the cases.

A continuous star formation history of the stellar ionizing population
has been assumed in all the objects of the sample. This
is supported by the simultaneous presence of very massive O
and WR stars, along with the relatively low measured values of EW(H$\beta$), once
corrected for underlying contribution and dust absorption.
The ages of the star formation bursts, the number of
ionizing photons as derived from the H$\alpha$ fluxes and the 
intensities of the most prominent emission lines were used to constrain photoionization models 
for each galaxy. The Starburst99 libraries were used to calculate the SED of the ionizing cluster 
in order to be consistent with the previous fitting of the stellar populations. Some dust is assumed 
to be present associated with the gas. The amount of dust in each model was derived in order 
to reproduce the observed EW(H$\beta$).
These derived values agree well with the main observed quantities in each object, with
the exception of the sizes of the ionized regions which result 5-10
times smaller in the models as compared to measurements from H$\alpha$ in the spatial profiles of the slits.

The observed electron temperatures and ionic abundances
are, in most cases, well reproduced by the models which implies that the direct method using collisional
emission lines provides robust answers 
if the thermal structure of the nebula is well traced. Therefore,
our models were able to reproduce the ionization structure
of the nebula without appealing to temperature fluctuations
or any other geometrical effects that can affect the
abundances derived from collisionally excited emission lines.
This result is consistent with the temperature measured from the Balmer
jump in some of the objects in our sample (H06). Only the
electron temperature and the ionic abundances of S$^+$/H$^+$ is not
well reproduced by our models, which could be consequence of the
presence of diffuse gas in these galaxies, not taken into
account in the models. This would also be consistent
with the small sizes predicted by the models as compared with H$\alpha$ spatial profiles
in the slits. 

Finally, regarding the hardness
of the radiation field, model predictions agree with observations
when the softness parameter $\eta$, which parametrizes the temperature of the ionizing radiation, is expressed in terms of emission line intensity ratios. However, there is a disagreement when ionic abundance ratios are used instead. This is probably 
caused by the overestimate of the electron temperature of S$^+$ by the models and the corresponding underestimate of the S$^+$/S$^{2+}$ abundance ratio rather than by existence of an additional
heating source in HII galaxies.

\section*{Acknowledgements}
This work has been supported by the CNRS-INSU (France) and its Programme National Galaxies and the project AYA2007- 67965-C03-02 and 03 of the Spanish National Plan for Astronomy and Astrophysics. Also, partial support from the Comunidad de Madrid under grant S0505/ESP/000237 (ASTROCAM) is acknowledged. EPM acknowledges financial support from the {\em Fundaci\'on Espa\~nola para la Ciencia y la Tecnolog\'\i a} and the Spanish {\em Ministerio de Innovaci\'on y Ciencia} for his postdoctoral position in LATT during two years and the awarding of a one-month grant 
to work at the Universidad Aut\'onoma de Madrid, to which EPM also thanks for its hospitality.
We also thank the anonymous referee whose constructive comments have helped to improve this work.

\appendix

\section{Total abundances and ionization correction factors}

The calculation of tailor-made photoionization models for each galaxy with a
reliable inner ionization structure based on the measurement of various electron
temperatures allows a complete description of the total abundances of the
main elements present in the gas. With this information it is possible to compare 
the ionization fractions obtained from the models with the most
widely used ionization correction factors (ICFs), necessary to derive
the total abundances of certain elements for which there are no lines in the optical range for all the main ionization stages.
Generally, for an ionic species $X^{+i}$,
whose abundance has been derived from forbidden collisional lines, we define
the ICF for that ion in the following manner:

\begin{equation}
\frac{X}{H} = ICF(X^{+i}) \cdot \frac{X^{+i}}{H^+}
\end{equation}

In Table \ref{icfs} we list the total abundances and the ICFs obtained from the models
for N, S, Ne, Ar and Fe.


\begin{table*}
\begin{minipage}{180mm}
\begin{center}
\caption{Ionization correction factors and total abundances as derived from photoionization models of the main elements with emission lines in the optical spectral range of the studied HII galaxies.}
\begin{tabular}{lcccccccccc}
\hline

  & J0021 & J0032 & J1455 & J1509 & J1528 & J1540 & J1616 & J1624 & J1657 & J1729 \\ 
\hline
12+log(O/H) & 7.99 & 8.01 & 7.94 & 8.18 & 8.18 & 8.15 & 8.00 & 8.04 & 8.01 & 8.12\\
\\
ICF(N$^+$) & 4.17 & 4.79 & 6.61 & 5.13 & 3.80 & 3.47 & 9.78 & 5.13 & 3.39 & 5.37\\
12+log(N/H) &  7.13 & 6.78 & 6.76 & 7.01 & 7.00 & 6.90 & 6.62 & 6.71 & 6.76 & 7.16\\
\\
ICF(S$^+$+S$^{2+}$)  & 1.11 & 1.14 & 1.28 & 1.12 & 1.10 & 1.04 & 1.37 & 1.16 & 1.07 & 1.19\\
12+log(S/H) & 6.29 & 6.40 & 6.23 & 6.57 & 6.41 & 6.54 & 6.33 & 6.30 & 6.34 & 6.41\\
\\
ICF(Ne$^{2+}$) & 1.20 & 1.17 & 1.10 & 1.20 & 1.23 & 1.29 & 1.07 & 1.15 & 1.26 & 1.15\\
12+log(Ne/H) & 7.24 & 7.24 & 7.16 & 7.36 & 7.43 & 7.22 & 7.17 & 7.22 & 7.19 & 7.37\\
\\
ICF(Ar$^{2+}$) & 1.15 & 1.17 & 1.23 & 1.15 & 1.15 & 1.10 & 1.29 & 1.17 & 1.15 & 1.17\\
ICF(Ar$^{2+}$+Ar$^{3+}$) & 1.05 & 1.06 & 1.03 & 1.04 & 1.07 & 1.06 & 1.02 & 1.04 & 1.08 & 1.04\\
12+log(Ar/H) & 5.58 & 5.69 & 5.58 & 5.87 & 5.78 & 5.82 & 5.64 & 5.68 & 5.57 & 5.79\\
\\
ICF(Fe$^{2+}$) & 2.69 & 2.88 & 3.89 & 2.95 & 2.57 & 2.14 & 5.13 & 3.09 & 2.40 & 3.24\\
12+log(Fe/H) & 5.82 & 5.80 & 5.86 & 5.92 & 5.99 & 5.73 & 6.15 & 5.77 & 5.79 & 5.88\\

\hline
\label{icfs}
\end{tabular}
\end{center}
\end{minipage}
\end{table*}


\begin{figure*}
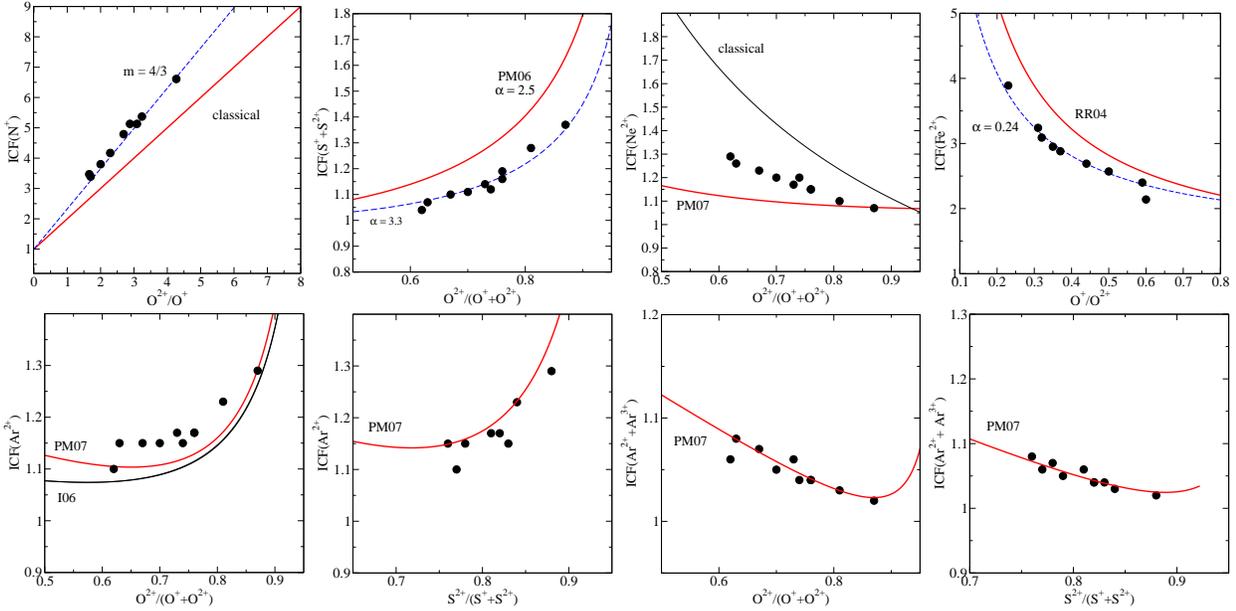

\begin{minipage}{170mm}
\centerline{
\psfig{figure=mn-ghiimod.fga1a.eps,width=4cm,clip=}
\psfig{figure=mn-ghiimod.fga1b.eps,width=4cm,clip=}
\psfig{figure=mn-ghiimod.fga1c.eps,width=4cm,clip=}
\psfig{figure=mn-ghiimod.fga1d.eps,width=4cm,clip=}}
\centerline{
\psfig{figure=mn-ghiimod.fg12e.eps,width=4cm,clip=}
\psfig{figure=mn-ghiimod.fg12f.eps,width=4cm,clip=}
\psfig{figure=mn-ghiimod.fg12g.eps,width=4cm,clip=}
\psfig{figure=mn-ghiimod.fg12h.eps,width=4cm,clip=}}
\caption{Comparison between different ionization correction factors (ICFs) derived for different elements from
the computed photoionization models (black circles) and some of the most used relations.}
\label{icf}
\end{minipage}
\end{figure*} 

The oxygen ionic abundance ratios, O$^{+}$/H$^{+}$ and O$^{2+}$/H$^{+}$, 
usually derived from the [O{\sc ii}]\,$\lambda\lambda$\,3727,29\,\AA\ and [O{\sc
    iii}] $\lambda\lambda$\,4959, 5007\,\AA\ lines are related to
the total abundance of oxygen assuming that:
\begin{equation}
\frac{O}{H}\,\approx\,\frac{O^++O^{2+}}{H^+}
\end{equation}

\noindent which is based on the assumption that the relative abundances of neutral hydrogen and oxygen
are similar. We have checked in our models that this assumption is valid within a 1\% of uncertainty.
As the bright emission lines of oxygen are well reproduced by our models, as also is the temperature sensitive line, the agreement between the total oxygen abundances
derived from observations and models agree fairly well with deviations, in most
cases, no larger than the reported observational errors.

The total abundance of nitrogen is obtained from the ionic abundance of  N$^{+}$,  derived from the
intensities of the $\lambda\lambda$\,6548, 6584\,\AA\ lines.
Then, a similar ionization structure for nitrogen and oxygen is assumed, according to the following expression:

\begin{equation}
\frac{N}{O}\,\approx\,\frac{N^+}{O^+}
\end{equation}

\noindent which, combined with equation (6), leads to the following ICF for nitrogen:

\begin{equation}
ICF(N^+) = 1+\frac{O^{2+}}{O^+}
\end{equation}

In the top left panel of Figure \ref{icf} we plot these ICFs as a function of the O$^+$/O$^{2+}$ ratio as compared to the best linear fit to the models. The slope of the fit is slightly higher (1.33), than the value derived using
the classical assumption. This is a consequence of the fact that the ionic fraction of N$^+$ is lower
than that of O$ ^+$ in these type of objects.

Regarding sulphur, we have derived S$^+$/H$^+$ abundances from  the fluxes of the [S{\sc ii}] emission lines at $\lambda\lambda$\,6717, 6731\,{\AA} and T([S{\sc ii}]) and S$^{2+}$/H$^+$ abundances from the fluxes of the near-IR  [S{\sc iii}]\,$\lambda\lambda$\,9069,9532\,{\AA}  lines and T([S{\sc iii}]).
The total sulphur abundance has been calculated using an ICF for S$^+$+S$^{2+}$ 
according to Barker's (1980) formula, which is based on the photoionization
models of Stasi\'nska (1978). We have written this formula  as
a function of the ratio O$^{2+}$/O instead of O$^+$/O in order to reduce the propagated error
for this quantity. 

 \begin{equation}
ICF(S^++S^{2+}) \approx \left[ 1-\left( \frac{O^{2+}}{O^++O^{2+}}
  \right)^\alpha\right]^{-1/\alpha} 
\end{equation}

\noindent where $\alpha$\,=\,2.5 provides the best fit to the scarce observational
data on S$^{3+}$ abundances (P\'erez-Montero et al.\ 2006, PM06 in the plot).  Nevertheless,
we can see in Figure \ref{icfs} that a value of $\alpha$ = 3.3 provides a better fit
to the results obtained from the photoionization models.

Neon is only visible in the spectra by means of the [Ne{\sc iii}] emission line at
$\lambda$3869\,{\AA}. The total abundance of neon has been
calculated using the following expression for the  ICF (P\'erez-Montero et al., 2007, PM07 in the plots):

\begin{equation}
ICF(Ne^{2+}) \approx 0.753 + 0.142\cdot  \frac{O^{2+}}{O^++O^{2+}} + 0.171 \cdot \frac{O^++O^{2+}}{O^{2+}}
\end{equation}

This formula takes into account the overestimate of
Ne/H in objects with low excitation, where the charge transfer between O$^{2+}$
and H$^0$ becomes important (Izotov et al., 2004, I04 in the plot). Nevertheless, the total abundances
obtained in our models are slightly higher than those derived with the use of this ICF.
 
The only available emission lines of argon in the optical spectra of ionized regions are those of [Ar{\sc iii}] at
7136 {\AA} and [Ar{\sc iv}] at 4740 {\AA}.
It is usually assumed that  T([Ar{\sc iii}])\,$\approx$\,T([S{\sc iii}])
(Garnett, 1992). The total abundance of Ar can be calculated  using the 
ICF(Ar$^{2+}$) derived from photo-ionization models by P\'erez-Montero et al. (2007, PM07 in the plots), 
in the case that only the emission line of  [Ar{\sc iii}] be detected
\[
ICF(Ar^{2+})\,=0.749+0.507\cdot\Big(1-\frac{O^{2+}}{O^++O^{2+}}\Big)+\]
\begin{equation}
+0.0604\cdot\Big(1-\frac{O^{2+}}{O^++O^{2+}}\Big)^{-1}
\end{equation}
In the same work, another ICF is proposed for the case in which [Ar{\sc iv}] is also measured: 

\[ICF(Ar^{2+}+Ar^{3+}) = 0.928 + 0.364\cdot (\Big(1-\frac{O^{2+}}{O^++O^{2+}}\Big)+\]
\begin{equation}
 + 0.006\cdot\Big(1-\frac{O^{2+}}{O^++O^{2+}}\Big)^{-1}
\end{equation}

These two ICFs can be expressed as well as a function of the 
ionic abundances of sulphur, S$^+$ and S$^{2+}$, when only red spectroscopic observations are
available:
\[
ICF(Ar^{2+})\,=0.596+0.967\cdot\Big(1-\frac{S^{2+}}{S^++S^{2+}}\Big)+\]
\begin{equation}
+0.077\cdot\Big(1-\frac{S^{2+}}{S^++S^{2+}}\Big)^{-1}
\end{equation}
\[ICF(Ar^{2+}+Ar^{3+}) = 0.870 + 0.695\cdot (\Big(1-\frac{S^{2+}}{S^++S^{2+}}\Big)+\]
\begin{equation}
 + 0.0086\cdot\Big(1-\frac{S^{2+}}{S^++S^{2+}}\Big)^{-1}
\end{equation}

In the bottom panels of Figure \ref{icf}, we can see that the agreement between these 
expressions and the ICFs found in our models is good.

Finally, iron abundances in the gas-phase can be calculated by correcting the
ionic abundances of Fe$^{2+}$, derived from the intensity of the emission
line of [Fe{\sc iii}] at 4658 {\AA}, using the following expression from Rodr\'\i guez \& Rubin
(2004, RR04 in the plot):
\begin{equation}
ICF(Fe^{2+}) = \left( \frac{O^+}{O^{2+}}\right)^\alpha \cdot \left[1+\frac{O^{2+}}{O^+}\right]
\end{equation}
\noindent with a value of $\alpha$ = 0.09. Nevertheless, the fit to the ICFs found
in our models gives a best value for $\alpha$ = 0.24, which leads to lower total abundances
of iron.

\end{document}